\documentclass[useAMS,usenatbib]{mn2e}
\usepackage{amsmath}
\usepackage{graphicx}
\usepackage{color}
\usepackage{afterpage}
\usepackage{float}
\usepackage{pdflscape}
\usepackage{amssymb}
\usepackage{enumitem}
\title[The IMF and SFL in the outer disk of M83]{The initial mass function in the extended ultraviolet disk of M83}
\author[S. M. Bruzzese et al.]{S.\ M.\ Bruzzese$^{1}$, David A.\ Thilker$^{2}$, G.\ R.\ Meurer$^{1}$, Luciana Bianchi$^{2}$, \newauthor{A.\ B.\ Watts$^{1}$}, 
A.\ M.\ N.\ Ferguson$^{3}$, A.\ Gil de Paz$^{4}$, B.\ Madore$^{5,6}$, \newauthor{D.\ Christopher Martin$^{7}$}, and R.\ Michael Rich$^{8}$\\
$^{1}$International Centre for Radio Astronomy Research, The University of Western Australia, Crawley, WA, Australia\\
$^{2}$Department of Physics and Astronomy, The Johns Hopkins University, Baltimore, MD, USA\\
$^{3}$Institute for Astronomy, Royal Observatory Edinburgh, University of Edinburgh, Blackford Hill, Edinburgh EH9 3HJ, United Kingdom\\
$^{4}$Departamento de F\'{\i}sica de la Tierra y Astrof\'{\i}sica, Universidad Complutense de Madrid, E-28040 Madrid, Spain\\
$^{5}$Department of Astronomy \&\ Astrophysics, University of Chicago, 5640 South Ellis Avenue, Chicago, IL 60637, USA\\
$^{6}$Observatories of the Carnegie Institution for Science, 813 Santa Barbara Street, Pasadena, CA 91101, USA\\
$^{7}$California Institute of Technology, Pasadena, CA 91125, USA\\
$^{8}$Department of Physics and Astronomy, UCLA, Los Angeles, California 90095-1547}

\newcommand{\HI}{\mbox{H\,{\sc i}}}

\newcommand{\Ha}{{H$\rm \alpha$}}
\newcommand{\Hb}{{H$\rm \beta$}}
\newcommand{\HII}{\mbox{H\,{\sc ii}}}

\newcommand{\mci}[1]{\multicolumn{1}{c}{#1}}

\begin{document}

\date{\today}

\pagerange{\pageref{firstpage}--\pageref{lastpage}} \pubyear{2016}

\maketitle

\label{firstpage}

\begin{abstract}
Using \textit{Hubble Space Telescope} ACS/WFC data we present the photometry and spatial distribution of resolved stellar populations of four fields within the extended ultraviolet disk (XUV disk) of M83. These observations show a clumpy distribution of main-sequence stars and a mostly smooth distribution of red giant branch stars. We constrain the upper-end of the initial mass function (IMF) in the outer disk using the detected population of main-sequence stars and an assumed constant star formation rate (SFR) over the last 300 Myr. By comparing the observed main-sequence luminosity function to simulations, we determine the best-fitting IMF to have a power law slope $\alpha=-2.35 \pm 0.3$ and an upper-mass limit $\rm M_{u}=25_{-3}^{+17} \, M_\odot$. This IMF is consistent with the observed \Ha\ emission, which we use to provide additional constraints on the IMF. We explore the influence of deviations from the constant SFR assumption, finding that our IMF conclusions are robust against all but strong recent variations in SFR, but these are excluded by causality arguments.  These results, along with our similar studies of other nearby galaxies, indicate that some XUV disks are deficient in high-mass stars compared to a Kroupa IMF. There are over one hundred galaxies within 5 Mpc, many already observed with HST, thus allowing a more comprehensive investigation of the IMF, and how it varies, using the techniques developed here.  
 
\end{abstract}


\section{Introduction}
\label{sec:intro}
A comprehensive understanding of star formation is essential to model and interpret the formation and evolution of galaxies. Star formation has been well studied in the bright, central regions of nearby galaxies, but is less well explored in their diffuse, low surface-brightness outer regions. Prior to the launch of the Galaxy Evolution Explorer (GALEX) satellite, it was commonly thought that outer disks of galaxies were stable and largely dormant \citep{kennicutt89,mk01}, although deep \Ha\ observations had shown the presence of some \HII\ regions, thus hinting that outer disks may be neither dormant nor pristine \citep{fwgh98,mk01}.  The GALEX Nearby Galaxy Survey revealed that outer disk star formation is common, occurring in nearly 30\%\ of nearby spiral galaxies \citep{gildepaz+05,Thilker:2005eh,thilker+05,thilker+07,zc07}.  The outer regions of disk galaxies have relatively low values for the stellar mass density, gas column densities, dust content, and metallicity compared to the central regions of galaxies \citep{GildePaz:2007ij, bigiel+10b, Barnes:2011eu}. Since outer disks contain much of the available supply of the interstellar medium \citep[ISM;][and references therein]{sfov08} they are crucial for understanding the current and future evolution of disk galaxies. 

The Initial Mass Function (hereafter IMF), the distribution in stellar mass of the stars that form in a young stellar population, is of critical importance to understand and model chemical enrichment and feedback processes in the ISM of these regions, the growth of galactic disks \citep{Larson:1976ic}, and how they depend on additional parameters such as metallicity \citep{Ostriker:2010dm,Krumholz:2013kk}.  It has been suggested that the IMF in low density environments may be deficient in high-mass stars \citep[e.g.][]{Elmegreen:2004jk,km08}, supported by observations which indicate that the upper-end of the IMF may vary with luminosity, surface brightness or star formation intensity \citep{hg08,meurer+09,lee+09hafuv,gunawardhana+11}. IMF variations, if they exist, could have enormous implications for the accuracy of determining the star formation rate (SFR) derived from indicators such as \Ha\ and UV emission, as well as star formation history (SFH) derived from colour-magnitude diagrams (CMDs). Because the outer disks of galaxies have low stellar and gas surface mass densities, and hence are well suited for characterising the IMF at the low density limit, they provide insight on how star formation varies with environment.

In this work, we study star formation in the extreme outer disk of M83, a nearby grand design spiral with an extended \HI\ disk \citep{bigiel+10a,heald+16}. This galaxy is a prototype extended UV (XUV) disk galaxy; its outer disk exhibits high-mass star formation readily detected in the UV by GALEX but faint at most other wavelengths \citep{thilker+07,Thilker:2005eh,thilker+05,gildepaz+05}.  M83 is also a prototype for galaxies with radially truncated star formation, evidenced by a sharp decline in azimuthally-averaged \Ha\ emission \citep{mk01,thilker+05}, which traces the most massive ionising stars. The discovery of its XUV disk \citep{thilker+05} illustrates that outer disk star formation in galaxies may be much more extensive than the earlier hints seen in \Ha.  M83's disk is also classified as extended in the UV as defined by the alternate technique developed by \citet{gkr10}.  Here, we use \textit{Hubble Space Telescope} images taken using the Advanced Camera for Surveys Wide Field Camera (ACS/WFC) to examine the resolved stellar populations of M83's XUV disk. These observations as well as \Ha\ data from Cerro Tololo Inter-American Observatory (CTIO) are used to place constraints on the upper-end IMF. 

This work follows on from that of \citet[][hereafter B15]{bruzzese+15} in which we developed a new technique to constrain the upper-end of the IMF and applied it to the outer disk of the gas-rich blue compact dwarf galaxy NGC 2915 using \textit{HST} observations of young high-mass main-sequence (MS) stars. We then applied the same technique to the outer disk of the dwarf irregular galaxy DDO~154 \citep{watts+18}.  Usually studies of resolved stellar populations assume a universal IMF \citep[e.g.\ that of][ also known as the Kroupa IMF]{kroupa01} and solve for the star formation history \citep[relatively recent examples include][]{gogarten+09,weisz+11a,weisz+13,annibali+13,lc13,meschin+14,lewis+15}. Instead, we adopt a plausible recent SFH of constant rate star formation and solve for the IMF. We also consider how sensitive our results are to recent departures from this steady evolution.  Using the same method as our previous studies allows us to make direct comparison between the derived IMF in each galaxy, providing insight as to how these relations depend on environmental conditions or galaxy type.  

This paper is organised as follows: In Section \ref{sec:obs_phot}, we present the data and its analysis.  We concentrate, particularly, on the HST/ACS observations and photometry of four outer disk fields in M83, but also discuss the \Ha\ data and their analysis. In Section \ref{sec:stellar_cont}, we present the colour-magnitude diagrams and stellar content of the resolved stellar populations of the observed outer disk fields, including the spatial distribution of stars in different evolutionary phases. In Section \ref{sec:IMF_con}, we outline the technique used to determine the best-fitting upper end IMF parameters and present our results. We do this primarily using the main-sequence luminosity function (MSLF) and then apply \Ha\ observations as additional constraints. Finally, our conclusions are presented in Section \ref{sec:concl}.

\begin{figure}
\centering
\includegraphics[width=85mm]{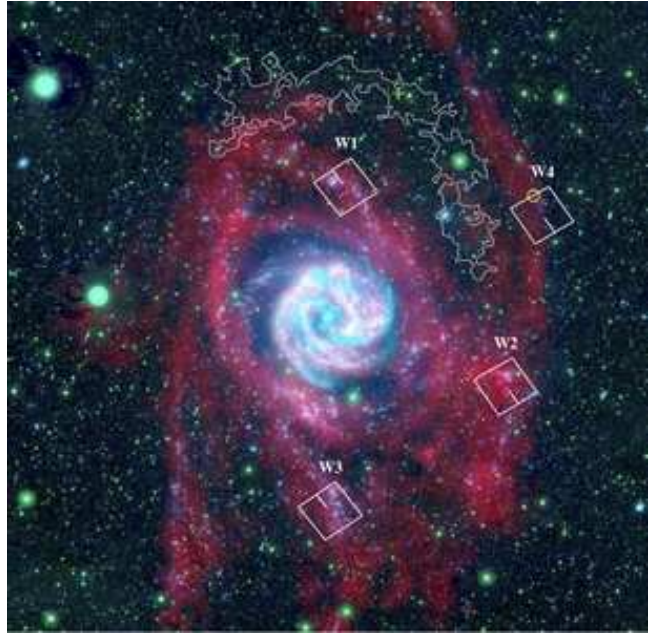}
\caption{Colour image of M83 composed of \HI\ Very Large Array (red), NUV GALEX (green), and FUV GALEX (blue) data.  This combination results in the strongly star forming central region to appear as a combination of white, cyan, and pink tones, while the outer disk is dominated by the \HI\ arms/filaments appearing in red, dotted in blue revealing the XUV disk star forming complexes.  The position of the four ACS/WFC fields are labelled and shown as white footprints. The short line segments in each footprint mark a portion of the divide between the two WFC chips and the position of the right edge of the colour images shown in Figures~\ref{fig:rgb_all}, \ref{fig:rgb_w1}, \ref{fig:rgb_w2}, and \ref{fig:rgb_w4}.  The position of KK208 is shown with a yellow cross.  The stream associated with KK208 is shown with a thin white contour, which is derived from the deep Spitzer IRAC observations of \citet{barnes+14}.  The galaxy dw1335--29, near the north-eastern edge of the W4 field, is marked with a yellow circle indicating its major axis half-light diameter \citep{carrillo+17}. The \HI\ data is from The \HI\ Nearby Galaxy Survey \citep{walter+08} and is compared to the UV data  in detail in \citet{bigiel+10a}.  (See published article for full resolution version of this figure).  \label{fig:M83_HLA}}
\end{figure}

\section{Observations and photometry}
\label{sec:obs_phot}
\begin{figure*}
\centering
\includegraphics[width=170mm, height=170mm]{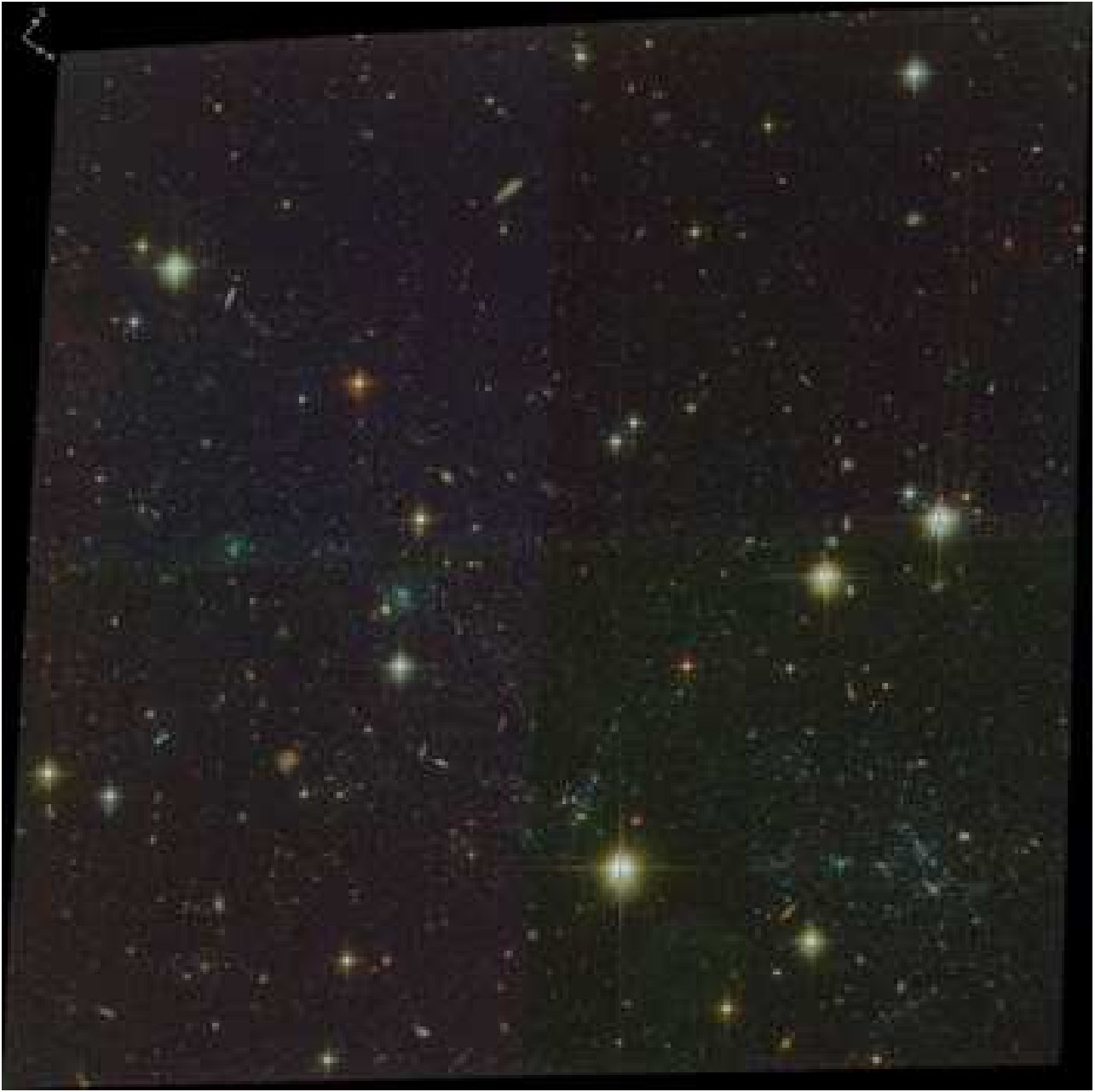}
\caption{Three colour $IVB$ image of the W3 field obtained with HST using ACS/WFC. The colour images for the other fields can be found in the Appendix. The dimensions of this image are 3.52$'$ in width and 3.58$'$ height (4.60 kpc $\times$ 4.69 kpc at our adopted distance) while the arrows indicating the cardinal directions at upper left are each 5$''$ (5.5 pc) long. (See published article for full resolution version of this figure). \label{fig:rgb_all}}
\end{figure*}

The primary data used here were taken using the Advanced Camera for
Surveys (ACS) Wide Field Camera (WFC) on the \textit{HST} (proposal ID:
10608, PI - Thilker) in 2006. Four ACS/WFC fields were chosen based on
the GALEX images to span a wide range in galacto-centric radius, UV
colour and morphology, and environment with respect to available \HI\
images. Other than requiring the images to be beyond the radius of the
strong \Ha\ edge seen in M83 \citep{mk01,thilker+05}, field selection
was not referenced to the presence or not of \Ha\ emission. Two
exposures were obtained of each field in the F435W, F606W, and F814W
filters (hereafter $B$, $V$ and $I$, respectively), with total exposure
times of 2522, 1190 and 890 seconds, respectively. The position of the
fields relative to M83 are shown in Figure \ref{fig:M83_HLA}. Here we
refer to the four fields as W1 through W4 (W for WFC).
Table~\ref{tab:field_info} specifies the location of the fields on the
sky and relative to the centre of M83, as well as information on how to
identify the the data in the HST data archive MAST\footnote{{\tt
    http://archive.stsci.edu/hst/}}. Figure \ref{fig:rgb_all} shows the
$IVB$ three colour drizzled image for the W3 field; the images for the
three other fields can be found in the Appendix. The \textit{HST} images
have a pixel size of 0.05 arcsec (1.06 pc at the adopted distance) and
each field covers an area of $\sim 4.6 \times 4.5$ kpc or
$20.4\, \rm kpc^{2}$. Table \ref{tab:M83_prop} lists the physical
properties of M83 we have adopted throughout this paper.

\begin{table}
\centering
\caption{The central position of each field and its deprojected distance $r$ from the centre of M83 at $\rmn{R.A.}=204.25375^\circ$, $\rmn{Dec.}=-29.865556^\circ$ (J2000). \label{tab:field_info}}
\begin{tabular}{|c|c|c|c|c|c|}
\hline 
Field & {Obs.} & {R.A.}  & {Decl.} & $\rm r$  & $\rm r$ \\
~     & set    & (J2000) & (J2000) & (arcmin) & (kpc) \\ 
\hline \hline
W1 & 53 & $204.230187^\circ$ & $-29.697383^\circ$ & 10.59 & 13.86 \\ 
W2 & 09 & $203.982992^\circ$ & $-29.956214^\circ$ & 16.40 & 21.47 \\ 
W3 & 62 & $204.253512^\circ$ & $-30.125311^\circ$ & 16.46 & 21.55 \\ 
W4 & 06 & $203.928788^\circ$ & $-29.730042^\circ$ & 18.96 & 24.82 \\ 
\hline 
\end{tabular} 
\end{table}

\begin{table}
\caption{Adopted M83 properties. \label{tab:M83_prop}}
\begin{tabular}{lcl}
\hline 
Property & Value  & Reference \\ 
\hline \hline
Z$^1$ & 0.006 ($\sim$ 1/3 $\rm Z_\odot$)  & \citet{Bresolin:2009im} \\
$E(B-V)^2$ & 0.06 mag &  \citet{Schlafly:2011iu} \\ 
$D^3$ & $4.5 $ Mpc & \citet{Karachentsev:2002fs} \\ 
$i^4$ & $25^{\circ}$ & \citet{Crosthwaite:2002cr} \\ 
PA$^5$ & 46$^{\circ}$ & \citet{Crosthwaite:2002cr} \\
\hline 
\end{tabular}
$^1$Metallicity in the outer disk over radii from 10 to 25 kpc. $^2$Foreground dust reddening. $^3$Distance. $^4$Inclination. $^5$Position angle of major axis measured from the North towards the East.
\end{table}

\subsection{Data reduction and measurement}
\label{sec:datared}
Initial image processing was done using \textsc{Calacs} to produce calibrated, flat-fielded FLT images  \citep{Hack:1999wl}. \textsc{AstroDrizzle} was used to combine the FLT images to produce a single, geometrically corrected, drizzled image per band. We performed stellar point source photometry on all four fields using the ACS module of the stellar photometry package \textsc{Dolphot} (v2.0), a modified version of \textsc{HSTphot} \citep{Dolphin:2000ib}. The photometry of all stars is derived by fitting pre-computed point spread functions to their image in each filter.  We use the drizzled \textit{B} image as the ``reference'' image; \textsc{Dolphot} is used to find sources in this image and then perform photometry simultaneously on all FLT images in all filters. We follow the processing steps outlined in the \textsc{Dolphot}/ACS User's Guide, adopting \textsc{Dolphot} parameters similar to those used by the ACS Nearby Galaxy Treasury (ANGST) team for crowded fields. The ANGST team found \textit{Force1}, \textit{RAper}, and \textit{FitSky} to have the strongest influence on photometry and we have set these parameters to their suggested values, as also done in B15. \textsc{Dolphot} provides the flux, position and quality parameters for each detected star in each filter in the Vega-MAG system and corrected for charge transfer efficiency loss according to \citet{Riess:2004wt}. We note that using the F435W filter as a reference will lead to an incomplete census of the red stellar population in comparison to the F606W or F814W filters. This is not a major concern for our analysis of the MS stars which are blue. A more complete study of the older stellar populations in this galaxy and other have been done by Galaxy Halos, Outer disks, Substructure, Thick disks and Star clusters (GHOSTS) team \citep{radsmi+11}.

To select stellar objects we applied the following quality cuts: $(S/N)_{1,2} \geq 4$, $|sharp_{1}+sharp_{2}| \leq 0.274 $, $crowd_{1}+crowd_{2} \leq 0.6$ and $flag_{1,2} \leq 2$, where $(S/N))$ is the signal to noise level of the detection, {\em sharp} is the sharpness parameter, {\em crowd} the crowding parameter, and {\em flag} is a quality flag.  All these quantities are produced by DOLPHOT, while the numbers refer to the filters. These cuts were done separately with \textit{B,I} or \textit{V,I} bands representing filters 1,2. The final catalogue is the union of the two separate cuts so that the survival in one set of cuts was sufficient for inclusion in our final catalogue. The {\em sharp\/} cuts were chosen to eliminate diffraction spikes, cosmic rays, blended stars, and background galaxies and the {\em crowd\/} cuts were chosen to eliminate stars with photometry significantly affected by crowding. These cuts are similar to those used by the ANGST team to produce the cleanest CMDs, minimising false stellar detections from extended sources and saturated pixels. In order to further minimise false detections from diffraction spikes and background galaxies we apply the masks created by the GHOSTS survey\footnote{Downloaded from https://archive.stsci.edu/pub/hlsp/ghosts/} for each field with minor modifications made to include some blue sources which were masked by the GHOSTS team. 

The cuts described above would exclude extended stellar clusters not resolved into stars, if present in our fields, and in principal could limit our inventory of the stars populating the MSLF.  However, we inspected the detections surviving our cuts over plotted on the multi-band HST images and find no evidence for clusters of this sort.  Furthermore, the resulting CMDs (Figures \ref{fig:M83_CMDs} and \ref{fig:cmd_all}) show that the detected source population does not contain anomalously bright sources (with respect to the stellar evolutionary tracks) that could be more properly interpreted as compact clusters.

\subsection{Artificial star tests}
\label{sec:artstar}
\begin{figure}
\centering
\includegraphics[width=85mm]{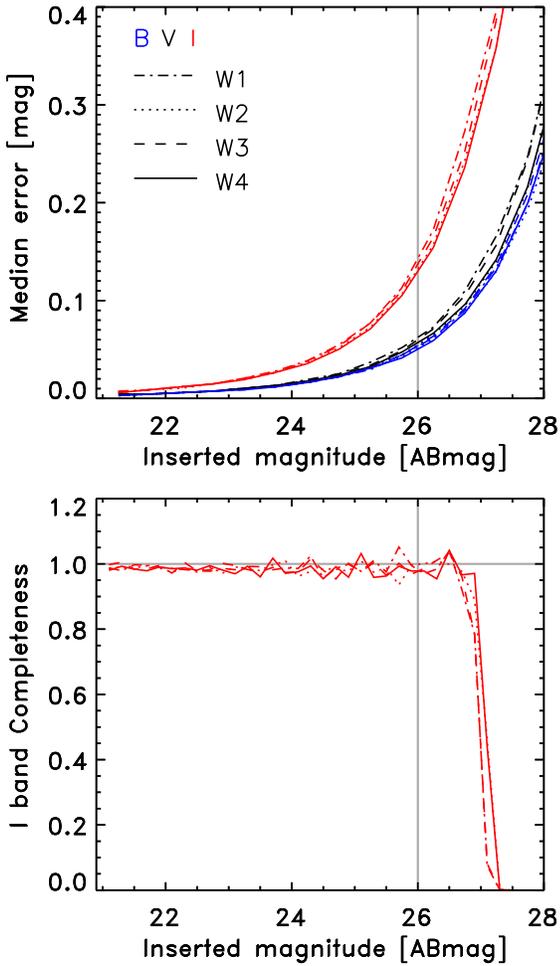} 
\caption{Top panel: Median photometric errors for each field. Bottom panel: completeness in the {\em I\/} band over the colour range encompassing the MS and BHeB stars.  The vertical gray lines mark the lower flux limit to our main-sequence selection box (see \ref{fig:M83_CMDs} below).  The horizontal gray line in the bottom panel marks unity completeness.  \label{fig:M83_AST_err}}
\end{figure}

To determine the photometric errors and completeness of the ACS/WFC data we generated $3 \times 10^5$ artificial stars per field distributed evenly across each of the fields, with colours and magnitudes comparable to the observations. We then re-ran \textsc{Dolphot} in artificial star mode to recover the photometry of the inserted artificial stars. The same photometric cuts that were applied to the real photometry were also applied to the artificial stars to create a catalogue of found artificial stars. The median photometric error is computed using Gaussian statistics as the median absolute deviation between the inserted and recovered magnitude for all recovered stars. The median photometric errors for each field and each filter are shown in the upper panel of Figure \ref{fig:M83_AST_err}. Completeness is determined as the fraction of inserted stars to recovered stars, in binned regions of colour and magnitude (with bin dimensions of 0.25, 0.5 mag, respectively). The 60\%\ completeness limit is shown for each CMD in red in Fig.~\ref{fig:M83_CMDs}. In order to improve statistics over the colour range most relevant to the science presented here, we also calculated completeness over the colour range in $B-I = -1$ to 1 (i.e. covering the MS and blue Helium burning stars) in bins of 0.2 mag in the $I$ band. The completeness as a function of $I$ in these bins is shown in the bottom panel of Fig.~\ref{fig:M83_AST_err}. These tests show that the main features of the CMD suffer minimal incompleteness ($\lesssim 5$\%).

\subsection{\Ha\ data}
\label{sec:Halpha}
In addition to the optical HST data we use a wide field of view \Ha\ image, which covers the full extent of the XUV disk to analyse the ionising stellar populations. The \Ha\ data were obtained using the CTIO Michigan Curtis Schmidt telescope in May 2001 as part of the Survey for Ionization in Neutral Gas Galaxies \citep[SINGG;][]{meurer+06}. Three exposures with a total exposure time of 1800 seconds were obtained in both the CTIO $R$ band filter and the 6568/30 narrow-band filter of the Magellanic Clouds Emission Line Survey \citep{smith+98}. The \Ha\ image is produced by subtracting the aligned and scaled $R$ band continuum from the narrow-band image \citep[for details see][]{meurer+06}. The \Ha\ image has a pixel size of 2.3 arcsec (50 pc) and a 1.5 degree field-of-view. The Schmidt Telescope images were taken in non-photometric conditions. We perform a boot-strapped flux calibration using CTIO 1.5 meter observations of the central regions of M83 obtained with the same filter set using the CTIO 1.5m telescope and reported in \citet{meurer+06}. 

\section{Stellar content}
\label{sec:stellar_cont}

\subsection{Colour-magnitude diagrams}
\label{sec:CMDs}
\begin{figure*}
\centering
\includegraphics[width=85mm]{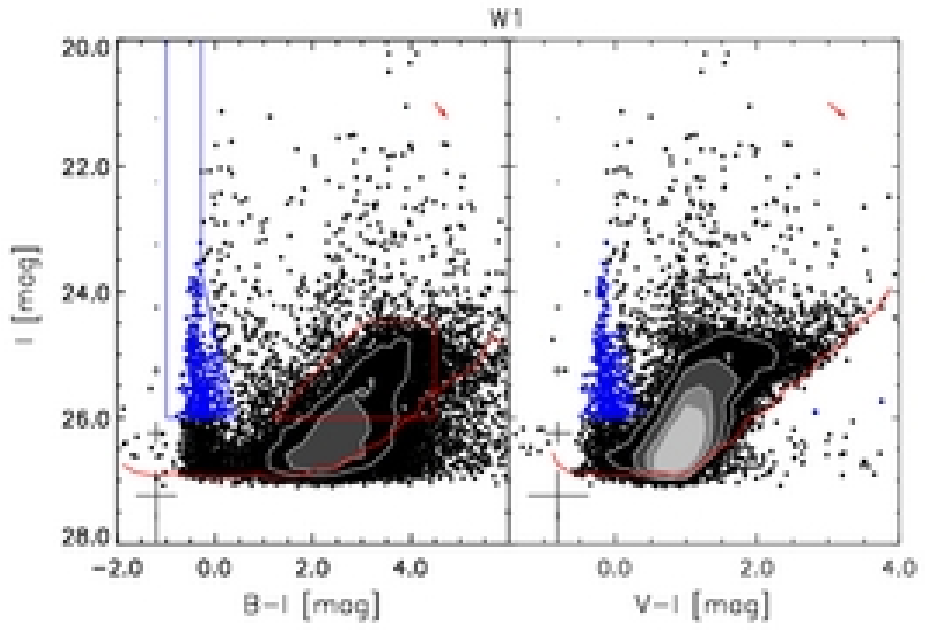}
\includegraphics[width=85mm]{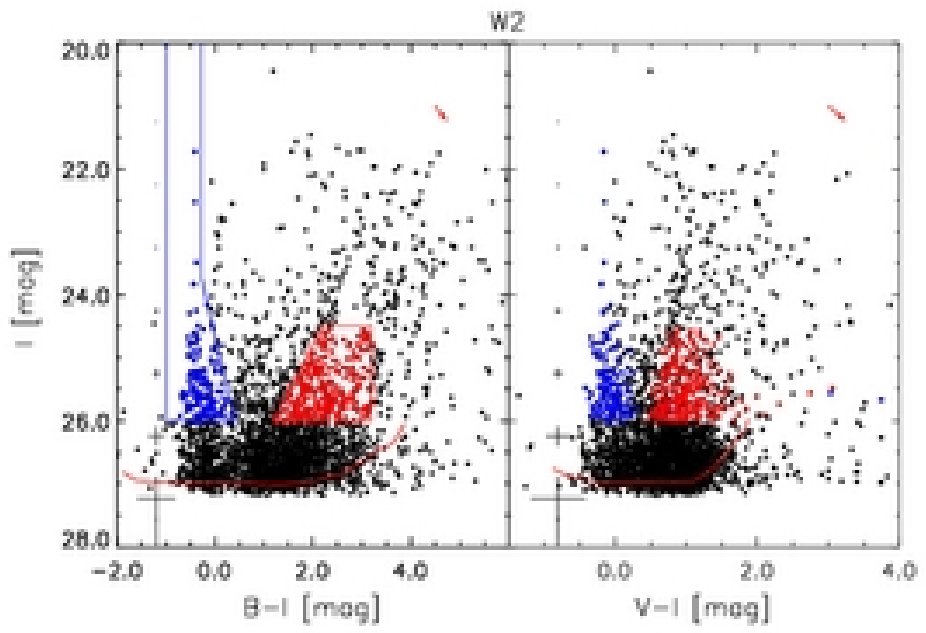}
\includegraphics[width=85mm]{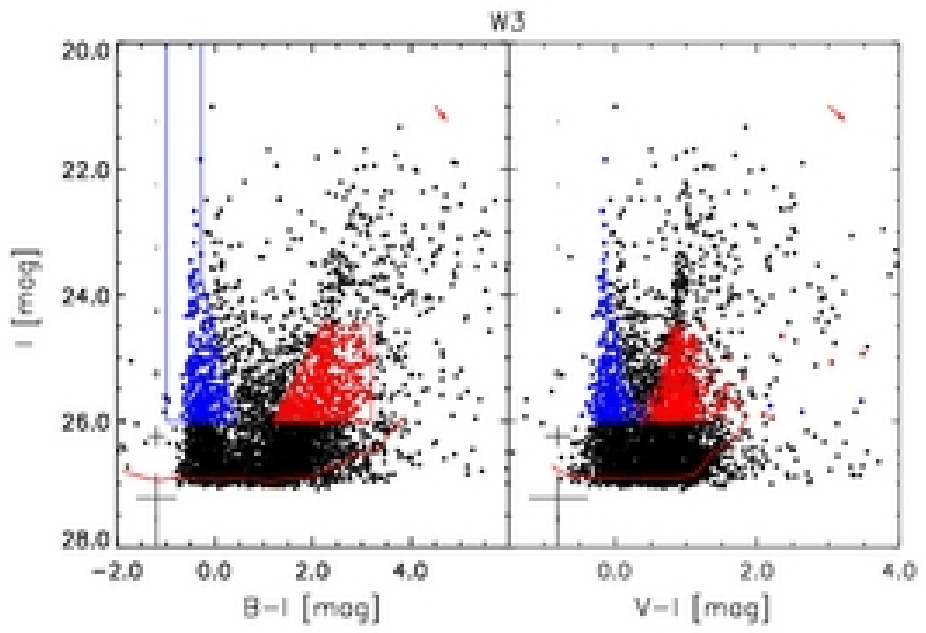}
\includegraphics[width=85mm]{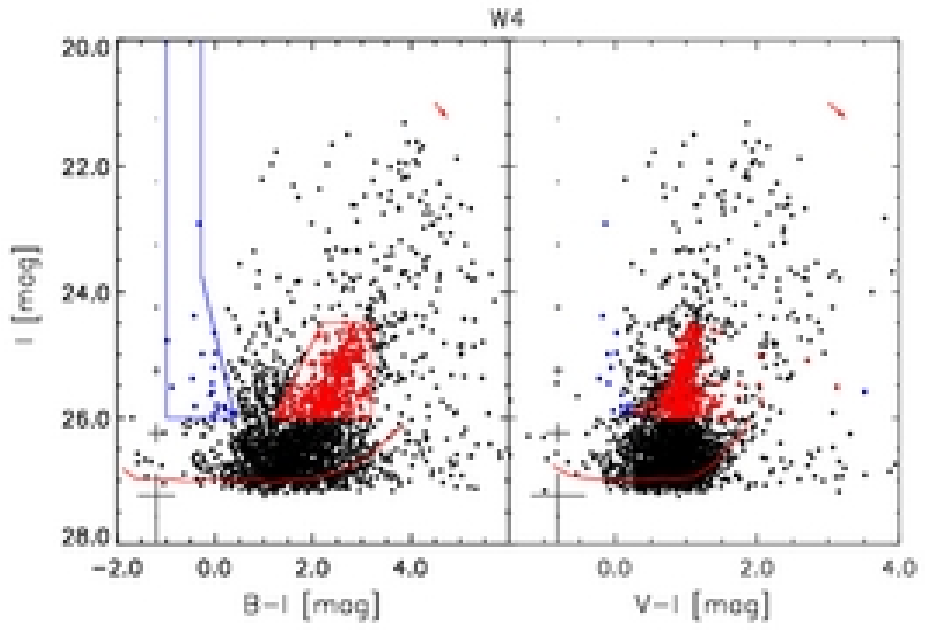}
\caption{CMDs of the ACS/WFC data in \textit{I} versus (\textit{B-I}) (left) and \textit{I} versus (\textit{V-I}) (right) for each of the individual HST fields. Polygon boxes, identical for each field, are shown in the (\textit{B-I}) CMDs, and are used to select stellar evolutionary phases; the blue polygon identifies MS stars, while the red polygon identifies RGB stars.  Stellar sources in these boxes are shown as blue and red dots respectively. Contours are used in the saturated portion of the colour-magnitude diagrams, calculated using bins 0.2 mag wide in both colour and magnitude, and contour levels at 50, 100, 200, and 300 stars per bin in each panel. Uncertainties derived from the artificial star tests are shown as error bars on the left side of the panels. The 60\%\ completeness limits are shown as thick red lines and the red arrows indicate assumed foreground dust extinction. \label{fig:M83_CMDs}}
\end{figure*}

\begin{figure*}
\centering
\includegraphics[height=65mm]{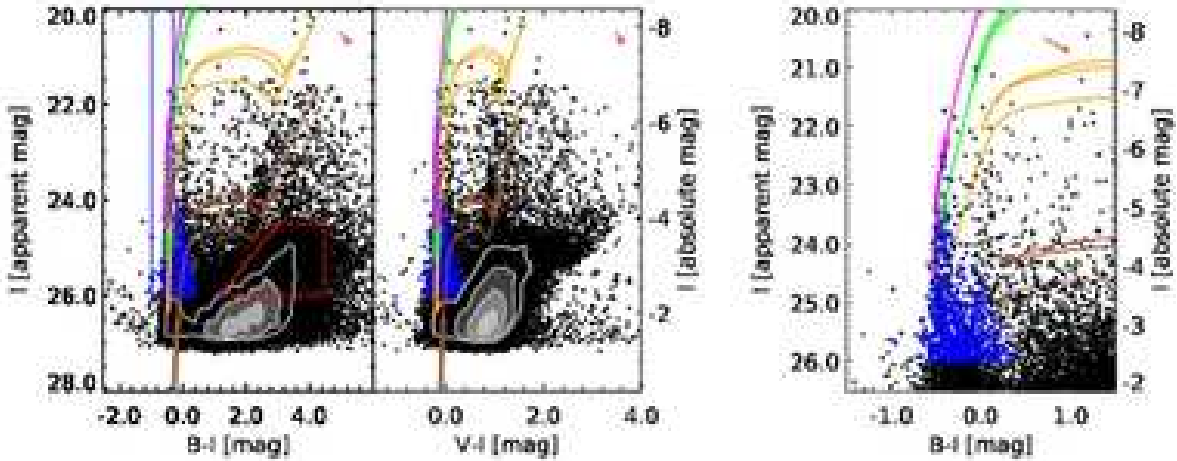}
\caption{CMDs combined from the data from all fields observed with ACS/WFC in \textit{I} versus (\textit{B-I}) (left) and \textit{I} versus (\textit{V-I}) (centre). Stellar sources identified from the photometry are shown as dots with blue representing stars in the MS selection polygon.  Contours are used in the saturated portion of the colour-magnitude diagrams, calculated using bins 0.2 mag wide in both colour and magnitude, and contour levels at 70, 120, and 200 stars per bin in the right panel and 100, 200, 400, and 600 stars per bin in the middle panel. PARSEC evolutionary tracks corresponding to 5, 12, 20, and 30 $M_\odot$ stars are shown in brown, orange, green, and magenta respectively, with thicker tracks corresponding to the MS phase. Polygons set in the \textit{B-I} CMD are used to identify stellar evolutionary phases: blue identifies MS stars; red identifies RGB stars. The apparent magnitude scale is plotted on the left hand vertical axis of the left and right hand panels, while the right hand vertical axis on the central and right panels shows the conversion to absolute magnitude. The right panel expands the \textit{I} versus (\textit{B-I}) CMD concentrating on the MS portion of the diagram.  The colouring of the points is the same as the other panels. The photometry of the stellar sources are plotted on top of the evolutionary tracks, and the contours are neglected in this version of the CMD, so as to more clearly illustrate the density of MS stars. \label{fig:cmd_all}}
\end{figure*}

Figure \ref{fig:M83_CMDs} shows the ($B-I$,$I$) and ($V-I$,$I$) CMDs for
each field separately. In Figure \ref{fig:cmd_all} we show the CMDs for
all fields combined, with PARSEC stellar evolutionary tracks
\citep{Tang:2014gt, Bressan:2012bx} corresponding to 5, 12, 20, and 30
$M_\odot$ stars over-plotted. The CMDs show the typical features of
composite stellar populations, such as the MS and RGB. In addition, the
blue and red Helium burning sequences can be discerned, especially in
the combined CMD. 

Polygon selection boxes are used to isolate the MS and RGB evolutionary
phases and are chosen to match their location on simulated CMDs (as
discussed in Section \ref{sec:sim_CMDs}). The exception is the W1 field in
which the RGB polygon is wider in colour to better match the
observations.  These selection boxes are shown in Figures
\ref{fig:M83_CMDs}, and \ref{fig:cmd_all}. The lowest (initial) mass for
stars found in the MS box is $\sim4\, M_\odot$, which have a MS lifetime
of $\sim 150$ Myr.  The right hand panel of Fig.~\ref{fig:cmd_all} shows
the bluer portion of the combined \textit{I} versus (\textit{B-I}) CMD
at an expanded scale, and with the evolutionary tracks plotted
``underneath'' the stars so as to highlight the MS stars.

Particular attention was paid to the MS selection box.  The objects found slightly to the red of the MS selection box with $I$ as bright as $\sim$21 ABmag are likely to be a combination of Blue Helium Burning (BHeB) stars, stars that have just turned off the MS, dust reddened stars, multiple star systems and chance blends. Other effects that have been shown to widen the MS and shift objects to the red include mass transfer in binary system (blue stragglers) and stellar mergers \citep[e.g.][and references therein]{lgdm2017a,lgdm2017b,bdsb2019}. Our MS selection box is designed to avoid these objects.  The combined CMDs shown in Fig.~\ref{fig:cmd_all} shows there is a distinct sequence to the blue of the MS at $B-I \approx 0$, $V-I \approx 0$, and that it corresponds well to the blueward most excursion of the Helium burning loop in the PARSEC evolutionary tracks, indicating that this is dominated by the BHeB sequence.  Since we employ the $I$ band as the luminosity measure in our CMD diagrams the post-MS stars are significantly {\em brighter\/}, than their MS counterparts of the same initial mass, even though their bolometric luminosities are more similar.  This is clearly shown in the evolutionary tracks plotted in Fig.~\ref{fig:cmd_all}. Likewise, at a given $I$ luminosity the strength of the BHeB is boosted via the IMF and the increased duration of this phase with lower mass.  This effect is enhanced with a steeper IMF slope as we find in this study.  HST based optical CMDs of nearby galaxies show a wide range of relative strengths of the MS and BHeB sequences \citep[see e.g.][]{dalcanton+09}.  The location of the BHeB sequence and its strength depend on metallicity and which family of models is adopted \citep{cignoni+18}.  Examples similar to M83's outer disk where the BHeB is of at least comparable strength to the MS or reaches brighter $I$ band luminosity include UGC~4483, NGC~4068, NGC~4163, ESO~154-023 \citep{mcquinn+2010a}, NGC~7793 \citep{radburnsmith+12}, DDO~6 \citep{weisz+11a}, and M81 \citep{gogarten+09}.  

Because of these complications we have chosen a MS selection box that
avoids the BHeB sequence.  The MS box we adopt widens to the red at
lower luminosities to accommodate the red-ward drift of the MS as initial
stellar mass decreases, as well as the increasing errors.  Since the
separation between the MS and BHeB sequences also widens with decreasing
luminosity \citep{mcquinn+11,cignoni+18}, this widening does not
increase the contamination of BHeB stars within the MS box.  The box
used to select MS stars is better than 95\%\ complete according to our
artificial star tests (Fig. \ref{fig:M83_AST_err}, bottom panel) and the
RGB polygon suffers from minor incompleteness at low-luminosities. The
numbers of stars present in the two phases for each field are listed in
Table \ref{tab:phot_number}. We see that stars at both evolutionary
phases are present in all our fields.  

\begin{table}
\centering
\caption{Number of stellar sources in the final photometric catalogue for each field and in total for all fields. We specify the number of MS and RGB stars as defined by their position on the CMD (see Sec. \ref{sec:stellar_cont}) and their fractional contribution to the total number of stellar sources (in parentheses). \label{tab:phot_number}}
\begin{tabular}{crrr}
\hline
\mci{Field} & \mci{stellar sources} &  \mci{MS stars} &  \mci{RGB stars} \\
\hline \hline
W1  & 14477 & 296 (0.02) & 3910 (0.27) \\
W2  & 2936  & 159 (0.05) &  284 (0.10)\\
W3  & 3868  & 314 (0.08) &  474 (0.08) \\
W4  & 2535  &  20 (0.01) &  289 (0.11)\\
All & 23816 & 789 (0.03) & 5270 (0.22) \\
\hline
\end{tabular}
\end{table}

\begin{figure*}
\centering
\includegraphics[width=85mm, height=85mm]{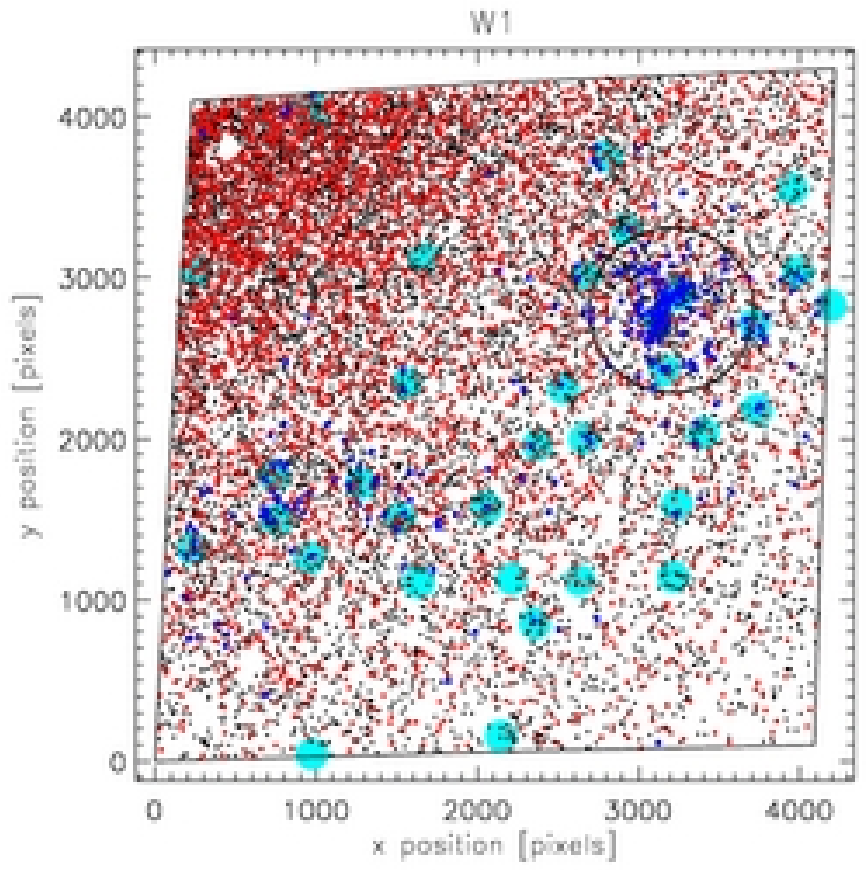} 
\includegraphics[width=85mm, height=85mm]{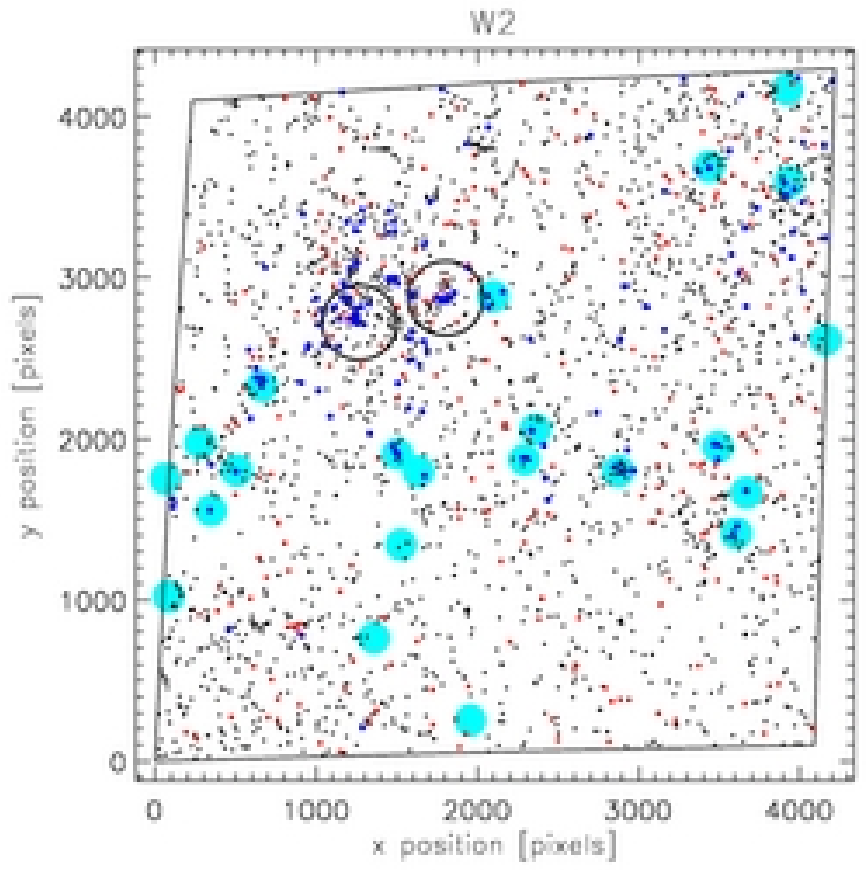} 
\includegraphics[width=85mm, height=85mm]{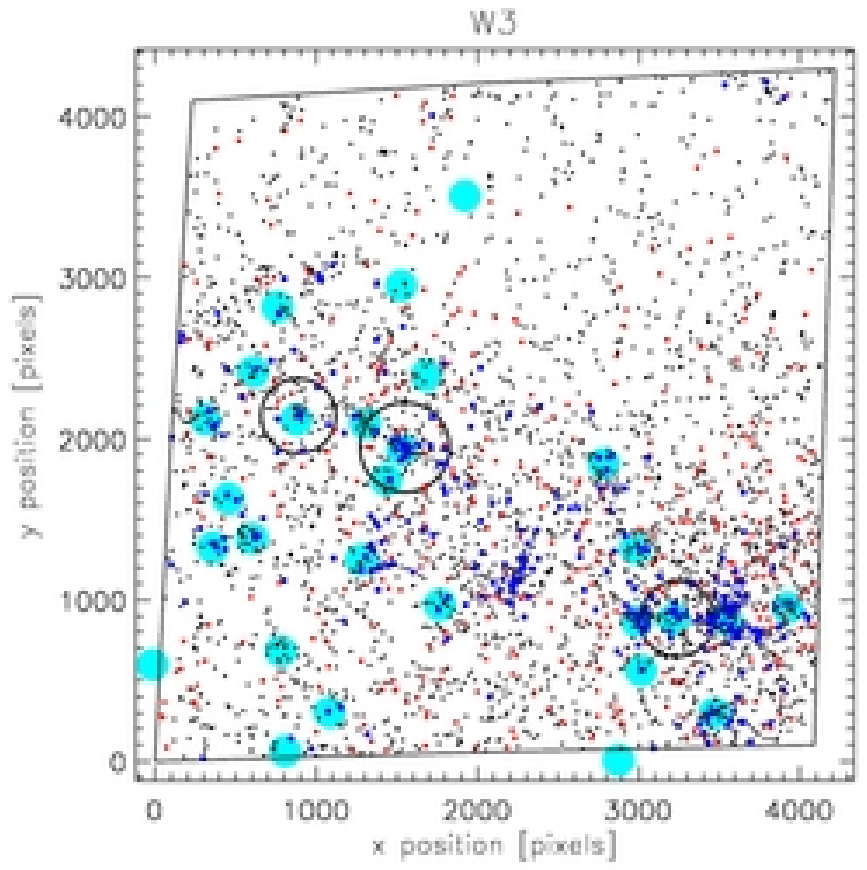}
\includegraphics[width=85mm, height=85mm]{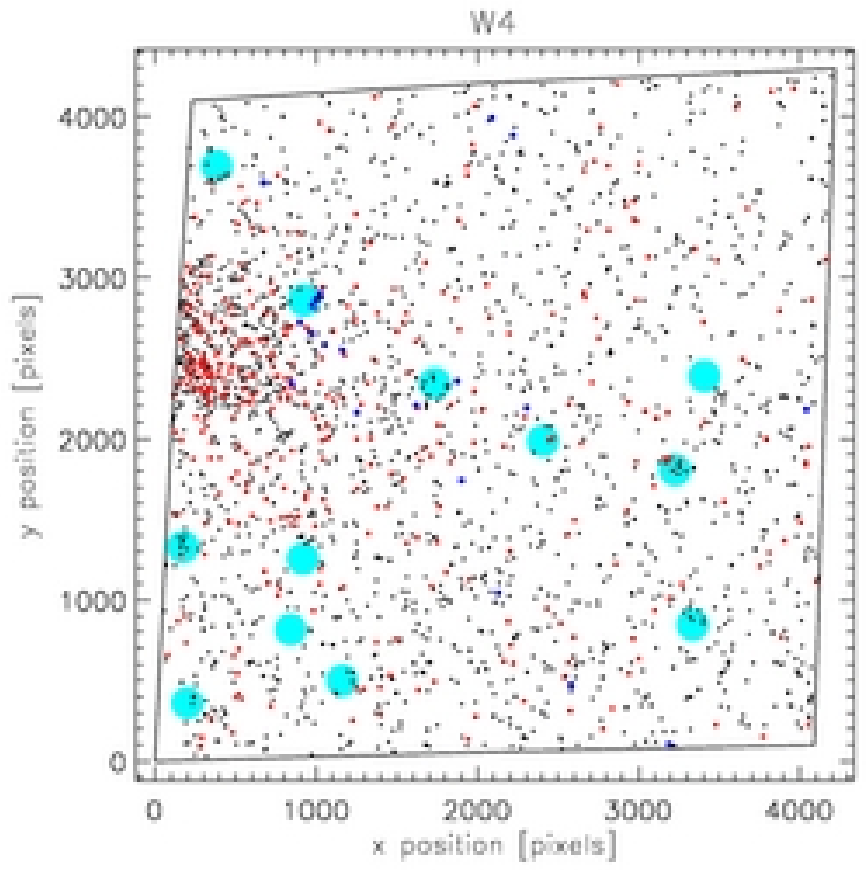}
\caption{Distribution of stars for each field. MS stars are marked in blue, RGB stars in red, and other stars in black. The selection boxes for the MS and RGB stars are shown in Fig.~\ref{fig:M83_CMDs}. The large black trapezoid shows the area covered by the WFC CCDs. Black circles outline the apertures used to measure the \Ha\ photometry of the \HII\ regions analysed in Section \ref{sec:HII_regs}. Cyan dots (5$''$ in radius) mark the positions of the FUV selected clusters studied by \citet{koda+12}. 
\label{fig:M83_spat_dist}}
\end{figure*}

Table \ref{tab:phot_number} lists the number of MS, RGB and total stellar sources that make the final photometric catalogue for each field. In Figure \ref{fig:M83_spat_dist} we show the spatial distribution of the MS and RGB stars present in each field, as identified from their location in the CMDs. A brief description of the four fields, arranged in order of projected radius from the centre follows.  \textbf{\textit{W1:}} is closest to the centre of M83 and so has the most densely populated CMD including a very prominent MS and a strong RGB covering a wide colour range.  \textbf{\textit{W2}} and \textbf{\textit{W3}}, are at nearly identical radii.  They have prominent MS and RGB sequences which are more densely populated in W2 despite being slightly further from the centre. \textbf{\textit{W4:}} This field is the farthest from the centre of M83 yet shows recent star formation at very low levels with 20 stars occupying the MS selection box.  One of the UV bright sources in this field detected by GALEX, which partially prompted the field selection, is produced by a background galaxy (at ${\rm RA} = 203.92625^\circ$, ${\rm Dec} = -29.74833^\circ$).  Figure \ref{fig:M83_HLA} shows that all fields are coincident with \HI\ emission.  The overlap with \HI\ appears the most uniform for W2, while the remaining fields are traversed by apparent \HI\ arms.

The prominent RGB in all fields indicate stars which may have ages from $\sim 2$ Gyr \citep[e.g.\ the age of the open cluster NGC~6819;][]{kt04} to $\sim 13$ Gyr \citep[the age of the oldest globular clusters;][]{sdiw97,salaris+weiss02}; i.e. they are significantly older than the MS stars we observe in these fields. These RGB stars may be in the halo or the disk (thick or thin), hence we do not expect their distribution to necessarily match that of the MS stars. Indeed the ratio of MS to RGB stars varies widely from $\sim 0.08$ (fields W1 and W4) to $>0.5$ (fields W2 and W3). Of the fields observed, W1 at a projected distance of 13.9 kpc has the greatest overlap with the extended smooth low surface brightness stellar disk first imaged by \citet{mh97}.  The CMD of W1 also has the best expressed old red giant branch, with a clear RGB tip at $I \approx 24.5$.  The old RGB population also has a wide range in abundance which we estimate spans --1.5 to --0.7 dex in [Fe/H] and possibly more metal rich\footnote{The red edge of the RGB is near the completion limit for the W1 field, suggesting we are missing some metal rich stars. Since the density of stars in the RGB region of the CMD decreases to the red as well, it is less clear whether significant numbers of metal rich RGB stars exist beyond this metallicity range.  Likewise, the smattering of stars to the red of the RGB selection box of the other fields may trace a similar metal rich old stellar population as seen in W1, but in much decreased numbers.}.   We may infer that the stellar populations traced by the RGB reaches the somewhat high metallicity attained in the halos of luminous galaxies such as M31, M104, and Cen A \citep{mouhcine+05} and exhibits the wide range of metallicities also found in the stellar halo populations of these systems. 

The distribution of MS stars indicate the location of recent star formation ($< 150$ Myr ago). Most MS stars are grouped together in clumpy formations, while others are more diffusely distributed. In contrast, the RGB stars in all fields are more smoothly distributed than the MS stars.  In the inner most field, W1, the RGB stars are more densely distributed with a clear density gradient with galacto-centric radius. W4, the outermost field, has a clump of RGB stars on the left (eastern) edge of the ACS image. There is no enhancement corresponding to this clump in our CTIO Schmidt data (see Section \ref{sec:Halpha}).  This clump corresponds to the M83 companion galaxy dw1335--29 discovered by \citet{mjb15}.  A detailed analysis of this galaxy using, in part, some of the data presented here, is given by \citet{carrillo+17}. The dwarf galaxy candidate KK208 (discovered by \citealt{Karachentseva:1998ab} and further discussed by \citealt{Miller:2009hu}) also is projected close to W4.  However, as shown in Fig.~\ref{fig:M83_HLA} this irregular shaped source is likely a faint outer arm of M83 which does not overlap with our fields.

We model foreground Galactic stellar contamination using the population synthesis code TRILEGAL \citep{Girardi:2005bb} to estimate the distribution of Milky Way stars in the CMDs. To do this, we assume constant foreground Milky Way extinction ($E(B-V)=0.06$ assumed throughout this paper), a Kroupa IMF corrected for binaries (using the standard inputs: binary fraction = 0.3 with mass uniform ratios between 0.7 and 1), along with the standard inputs for the Milky Way for the position of each field. These simulations indicate that we expect very few foreground stars ($n=0.8$ per field, on average) to appear in our MS selection box, with mild contamination for the rest of the CMD. 

\renewcommand{\thefigure}{\arabic{figure}}

\subsection{\HII\ regions}
\label{sec:HII_regs}

\begin{table*}
\centering
\caption[Properties of \HII\ regions]{Properties of the \HII\ regions studied here. The columns are as follows: (1) Our adopted name; (2) the name given by \citet[][\HII-4 corresponds to multiple \HII\ regions in their study]{Bresolin:2009im}; (3) field identification; (4,5) \HII\ region position (J2000 equinox); (6) aperture radius in arcsec; (7,8) dust corrected \Ha\ fluxes in units of $10^{-15}\, {\rm erg\ cm^{-2}\ s^{-1}}$ - measured from our CTIO images (column 7) using the circular aperture specified by columns 4--6 and measured from the \citet{Bresolin:2009im} slit spectroscopy (column 8); (9) the log of the ionising photon rate required to produce the flux given in column (7) in units of photons s$^{-1}$; (10,11) the mass and spectral type of a single main-sequence O star that can provide the ionising flux listed in table (9) derived from Table 1 of \citet{Martins:2005bn} (\HII-4\ and \HII-5\ require multiple O stars to ionising them); (12) The SFR, in units of $10^{-5}\, \rm M_{\odot}\,yr^{-1}$, required to produce the \Ha\ flux given in column (7) using the calibration of \citet{meurer+09}, for an assumed \citet{salpeter55} IMF spanning the mass range 0.1 to 100 $M_\odot$; (13) The number of MS stars identified in our HST images.\label{tab:Halpha_data}}
\begin{tabular}{|l|c|c|c|c|c|c|c|c|c|c|c|r|}
\hline
\HII\  & B09    & Field & R.A.\ & Dec.   & $R_{app}$ & $F_{Ha,0}$ & $F_{Ha,0}$            & log $Q_0$  & Mass & Spec. & \Ha\ & $\rm N_{MS}$ \\  
name   & name    & name & (deg J2000) & (deg J2000) & ($''$) & this paper & B09 & {\small ($\rm s^{-1}$)} & ($\rm M_{\odot}$) & type  & SFR &  \\
\multicolumn{1}{c}{(1)} & (2) & (3) & (4) & (5) & (6) & (7) & (8) & (9) & (10) & (11) & (12) &\multicolumn{1}{c}{(13)} \\ \hline \hline
\HII-1 & 30       & W3 & 204.244375 & -30.099972 & 11.6 & $5.9\pm 0.34$ & 2.1 & 49.0 & 31-24 & O6-5.5 & 9.1 & 24 \\
\HII-2 & 27       & W3 & 204.242083 & -30.127639 & 13.9 & $14.1\pm 0.5$ & 1.7 & 49.4 & 37-46 & O5-4   & 22  & 56  \\
\HII-3 & 25       & W3 & 204.238333 & -30.136639 & 11.6 & $7.5\pm 0.34$ & 1.3 & 49.1 & 34    & O5.5   & 12  & 64 \\
\HII-4 & 26,31,33 & W1 & 204.245417 & -29.688111 & 25.5 & $32.5\pm 0.7$ & 3.5 & 49.7 & $>58$ & $<$O3  &  5.1 & 157 \\
\HII-5 & 1        & W2 & 203.986250 & -29.944167 & 11.6 & $28.1\pm 0.1$ & 4.9 & 49.7 & $>58$ & $<$O3 &  44  & 23 \\
\HII-6 & --       & W2 & 203.979542 & -29.949278 & 11.6 & $2.1\pm 0.21$ & --  & 48.6 & 24-26 & O7-7.5 &  3.3 & 36 \\
\hline
\end{tabular}
\end{table*}

\begin{figure*}
\centering
\includegraphics[width=85mm]{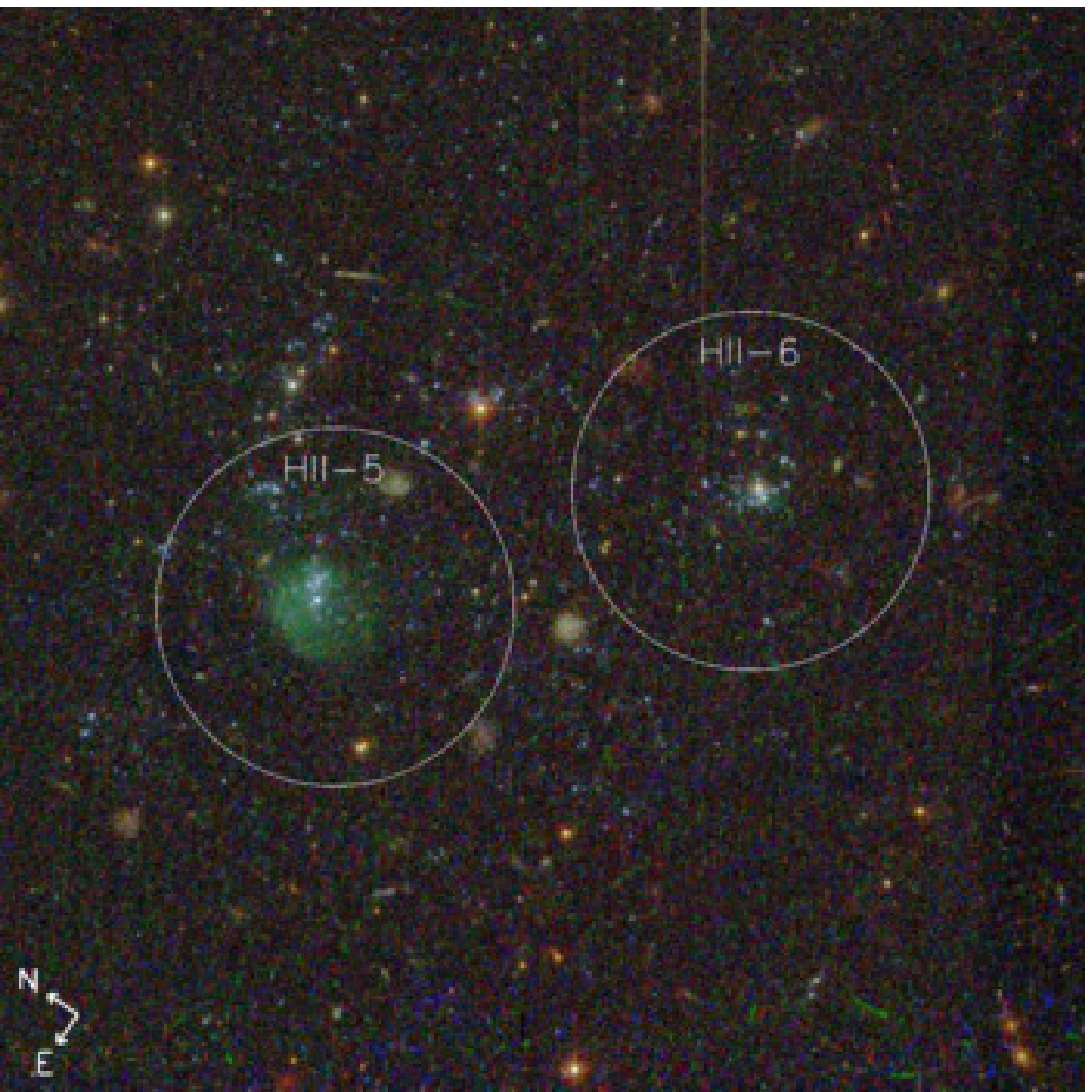}
\includegraphics[width=85mm]{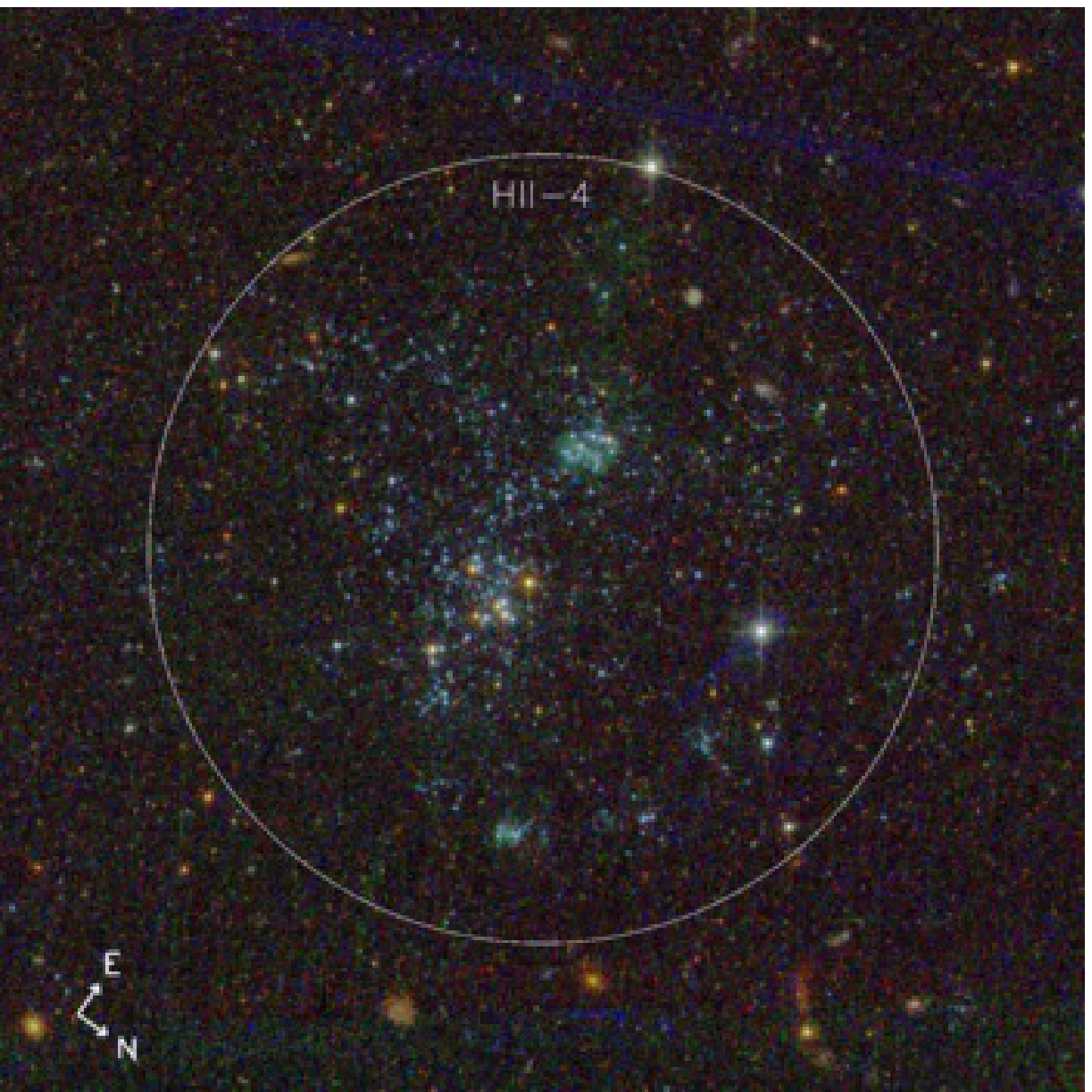}
\includegraphics[width=85mm]{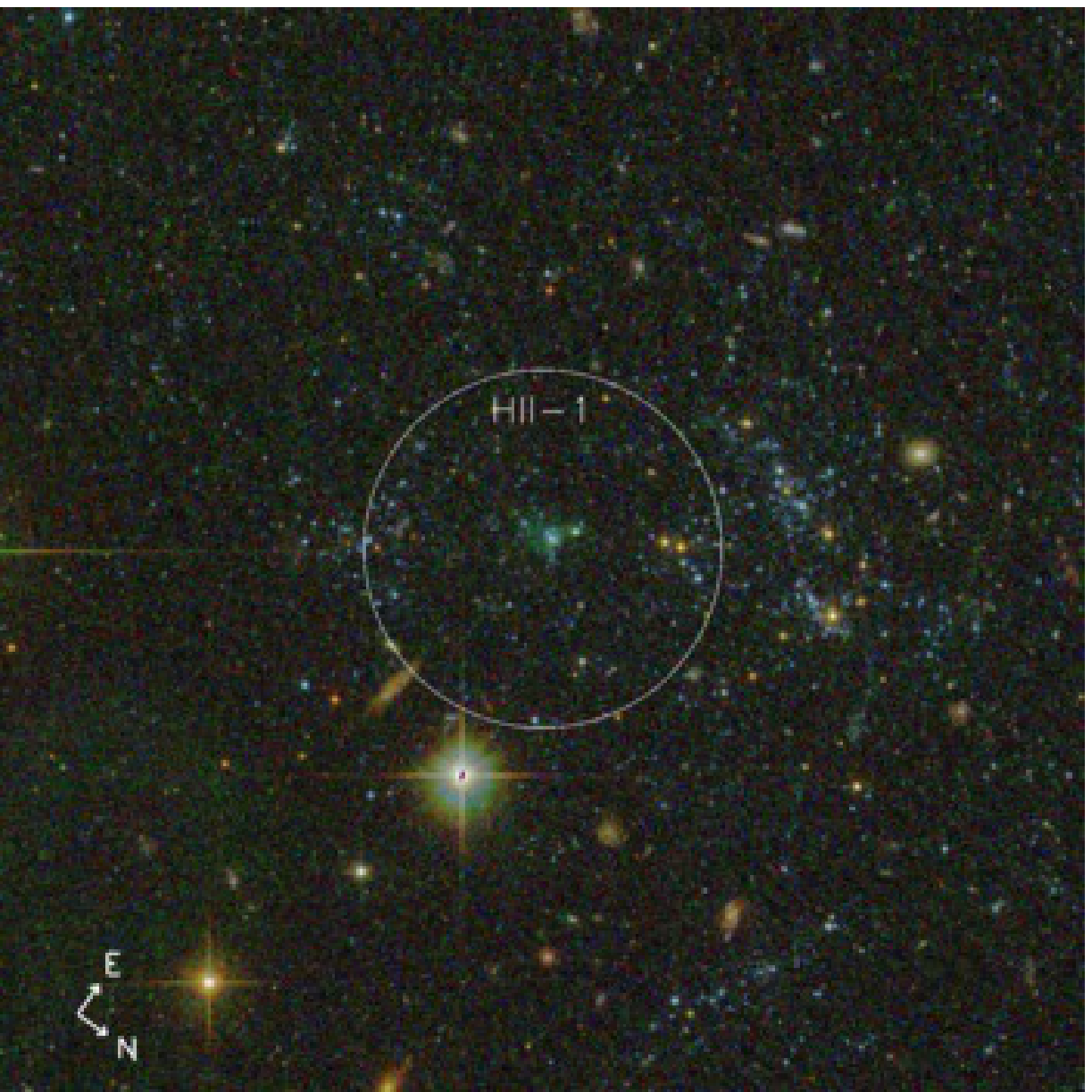}
\includegraphics[width=85mm]{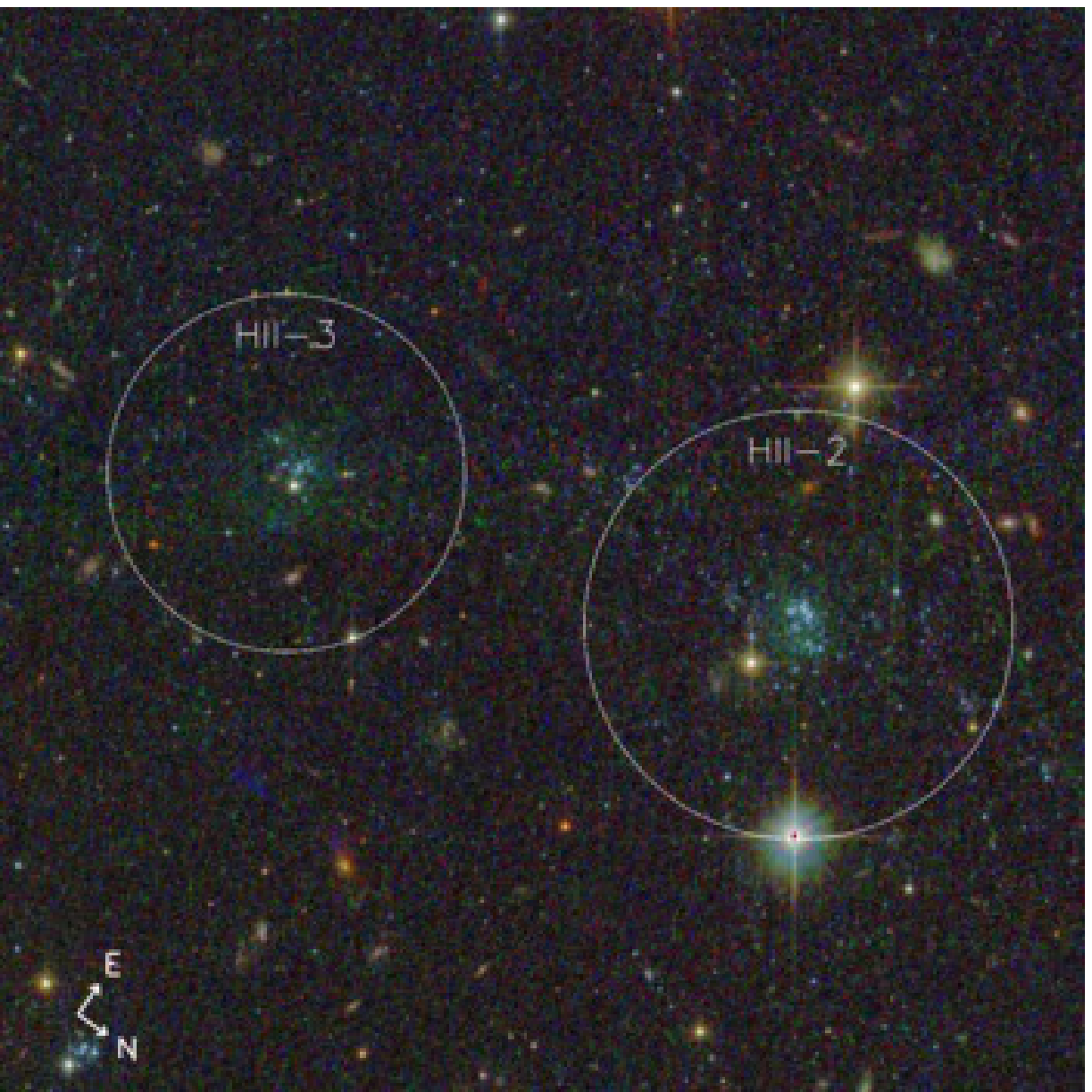}
\caption{The \HII\ regions identified in the CTIO \Ha\ data (Fig.~\ref{fig:Ha_img}) shown as three colour $IVB$ cut-outs from the HST images. The \HII\ regions are indicated by the circles, their radii and other properties are listed in Table \ref{tab:Halpha_data}. The name of each \HII\ region is indicated. Note the green diffuse emission seen within our adopted \HII\ region boundaries is due to the presence of emission lines including \Ha, \Hb, and [O{\sc iii}].  (See published article for full resolution version of this figure).  \label{fig:HII_BVI_close}}
\end{figure*}

We use the CTIO \Ha\ data described in Section \ref{sec:Halpha} to determine the properties of the \HII\ regions detected within the ACS/WFC fields. We use SExtractor \citep{ba96} to locate all \Ha\ bright regions and then manually check these against the $R$ band continuum image to verify the sources  (i.e. not a  poorly continuum-subtracted bright star, nor a noise spike). In Figure \ref{fig:HII_BVI_close} we show $IVB$ three colour cut-outs from the WFC images, which correspond to the \HII\ regions identified in the CTIO \Ha\ data. All of these regions coincide with groups of MS stars, some of which have an elongated appearance, indicating that there are neighbouring stars that have not been resolved. The stars associated with these \HII\ regions often have a greenish hue in the $IVB$ colour images, while others are embedded in a low surface-brightness green emission. This is likely due to contamination by emission lines such as H$ \beta $,  [O{\sc iii}] $\lambda \lambda$4959, 5007 \AA, \Ha, and [S{\sc ii}] $ \lambda \lambda$6716, 6731 \AA, which fall in the F606W $V$ band filter bandpass. Table \ref{tab:Halpha_data} lists the location of the \HII\ regions and other properties, as described below.

\begin{figure*}
\centering
\includegraphics[width=55mm]{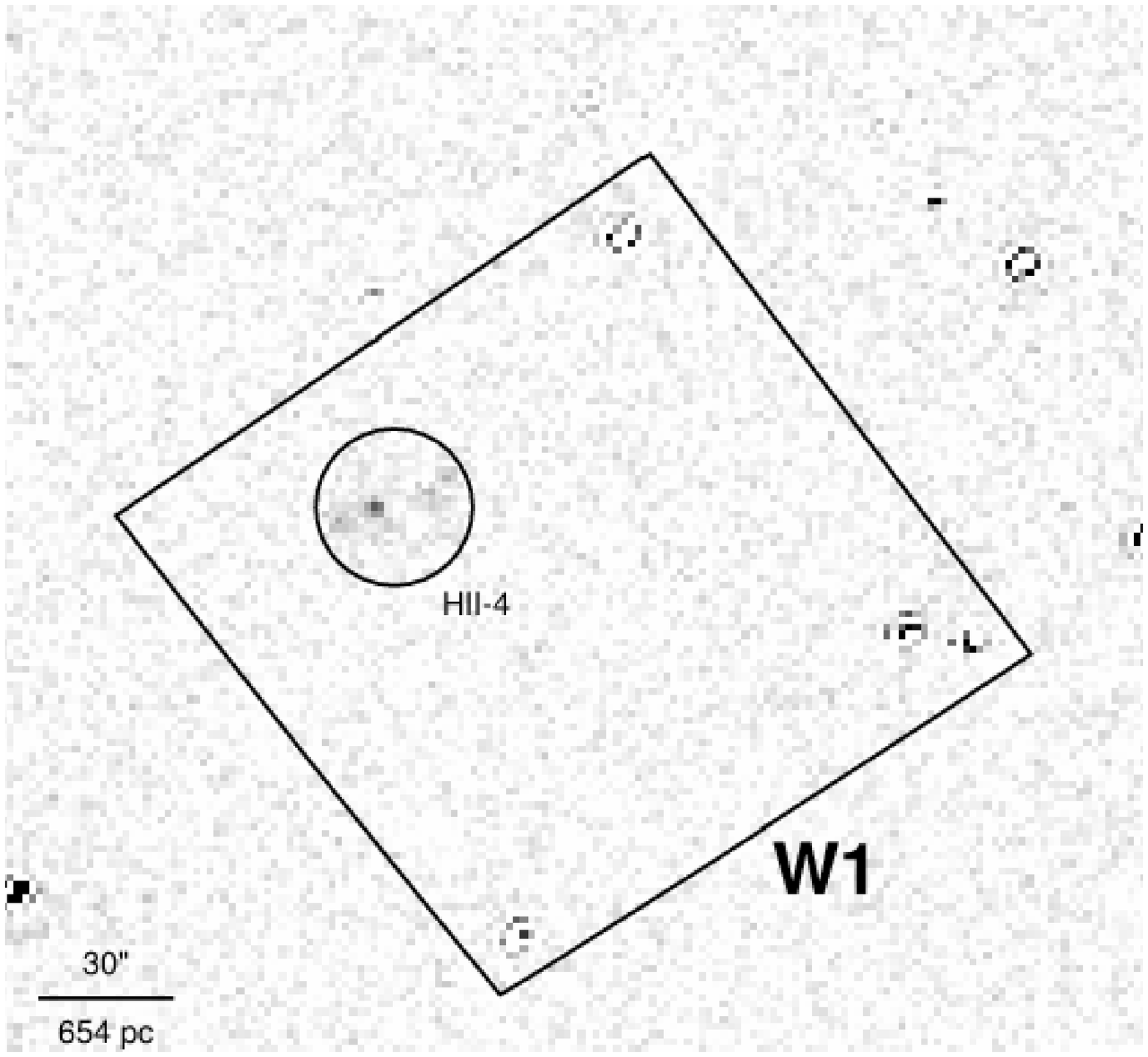} 
\includegraphics[width=55mm]{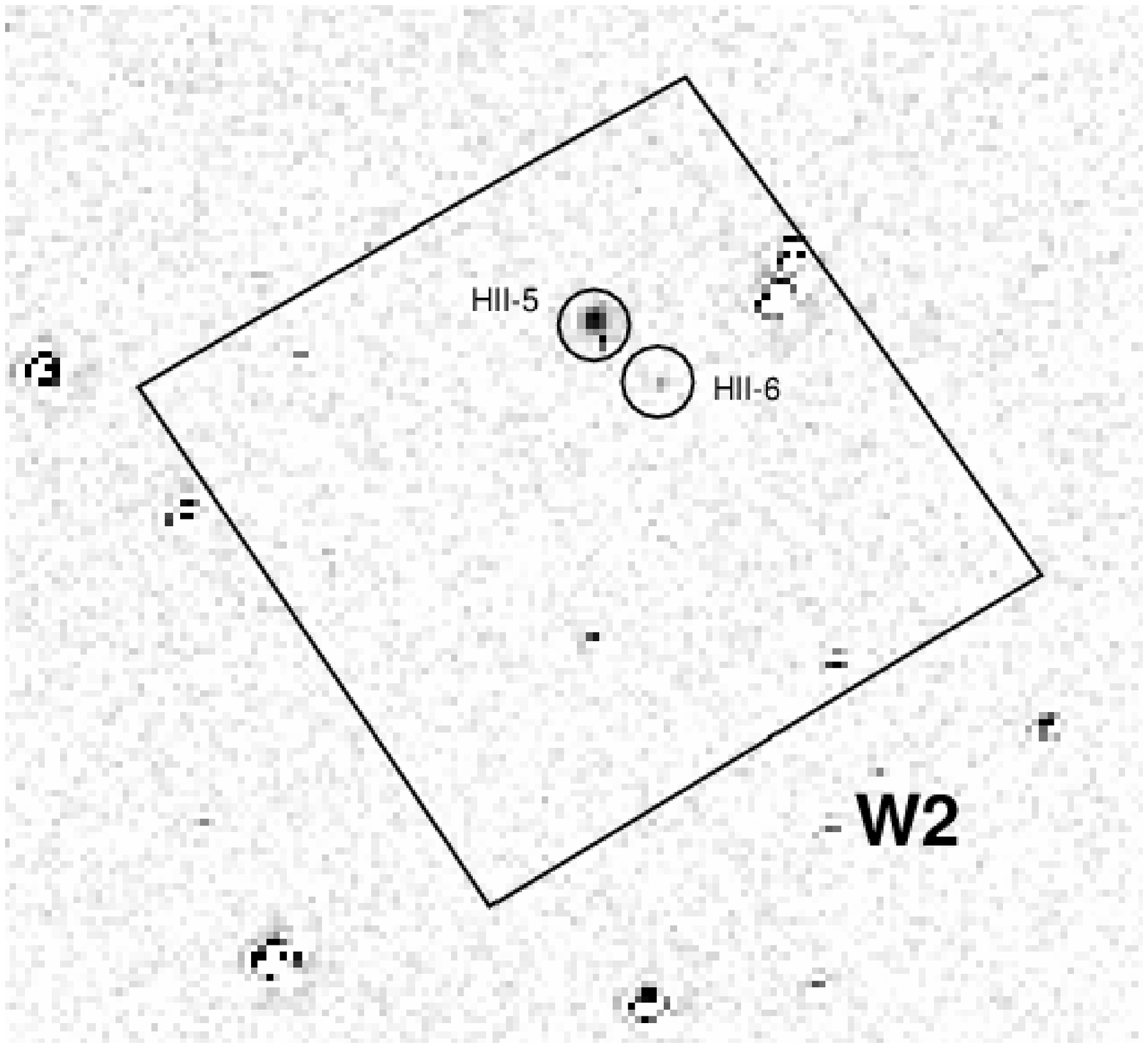} 
\includegraphics[width=55mm]{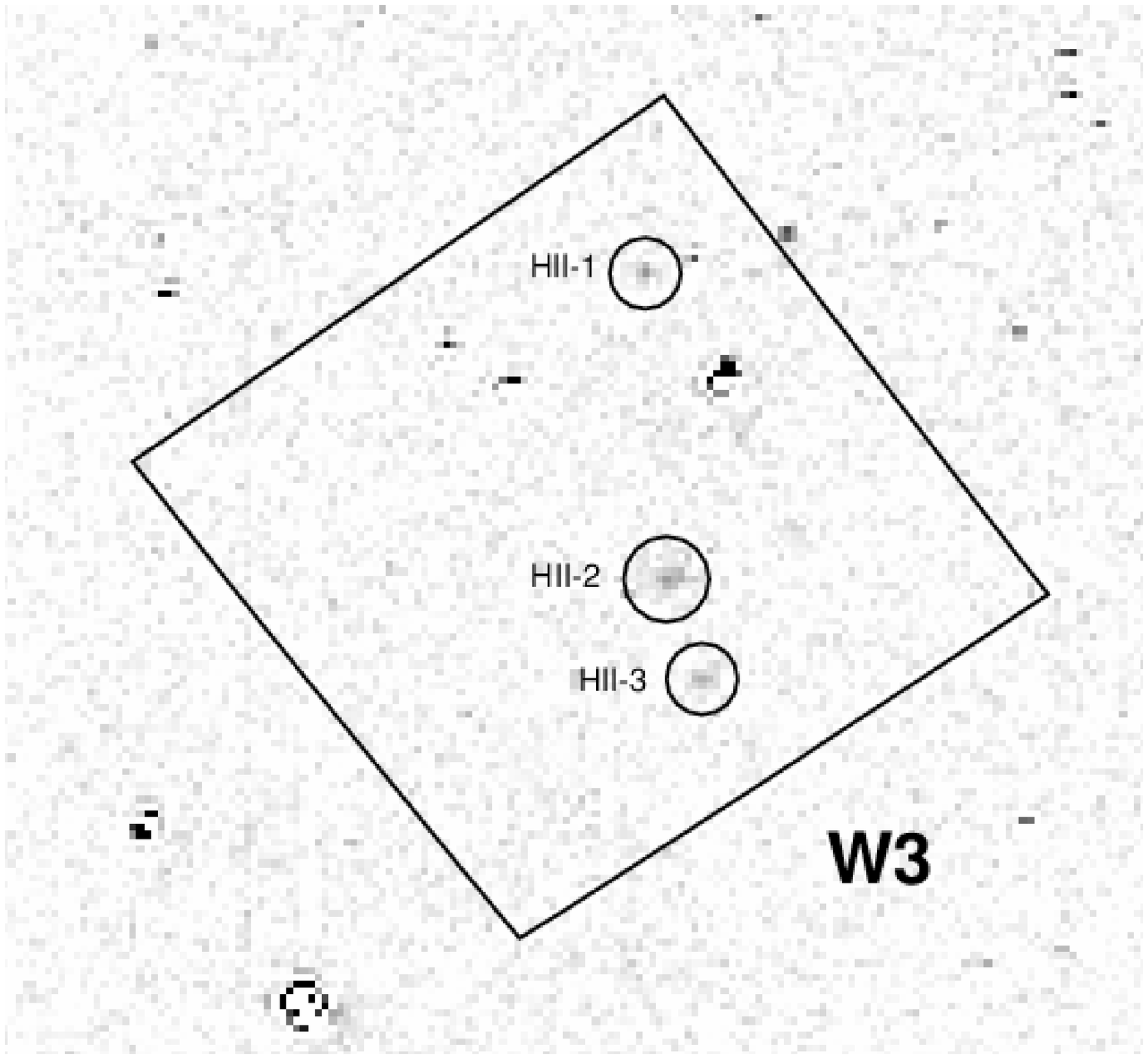} 
\caption[\Ha\ cutout images from our CTIO data]{Cutout \Ha\ images for each WFC field containing \HII\ regions, taken from our CTIO data. These show the HST footprint, and the apertures used to measure the \Ha\ flux as circles labelled with the \HII\ region number. The field name is given in the bottom right of each panel, while the image scale is shown in the panel for W1. 
} \label{fig:Ha_img}
\end{figure*}

We use aperture photometry to determine the \Ha\ flux from the CTIO images. All \HII\ regions appear to be a single sources at the resolution of the images, except for \HII-4, which appears as three extended, and partially blended, sources located within a 26 arcsec radius of the stated coordinates. We use a larger aperture to recover the combined flux of these.  In Figure \ref{fig:Ha_img} we show portions of the \Ha\ images along with the outlines of the four fields and the apertures used in the \Ha\ photometry. These images do not show any diffuse emission outside of the apertures. The cutouts in Fig.~\ref{fig:HII_BVI_close} show the ionised gas emission as a diffuse green glow in the HST data, and illustrate that significant diffuse emission is limited to our adopted apertures. 

Five out of the six \HII\ regions have been spectroscopically analysed by \citet{Bresolin:2009im}. We adopt their extinction corrections, and compare our \Ha\ flux measurements to theirs in Table \ref{tab:Halpha_data}. In all cases the \Ha\ flux we measure is larger than that of \citet{Bresolin:2009im}. This is due to aperture effects; they employed slitlets having a width of 1 arcsec, much smaller than the apertures we employ.  Our measurements also overestimate the \Ha\ flux due to contamination from the [N{\sc ii}]  $\lambda \lambda$6548, 6584 \AA\ lines, which fall within the \Ha\ filter passband. Using measurements of [N{\sc ii}] $\lambda$ 6583 \AA, from \citet{Bresolin:2009im} and multiplying by 1.3 to account for the [N{\sc ii}] $\lambda$ 6548 \AA\ line we estimate $\sim 20$\%\ contamination to the total \Ha\ flux from [N{\sc ii}] lines.

\HII\ regions indicate the presence of young ($<$10 Myr), high-mass ($\rm M_{\star}>15\,M_{\odot}$), O-type stars, which produce large amounts of ionising UV radiation.  We convert the \Ha\ luminosity of each \HII\ region to an ionising rate making the standard assumption that all ionising photons are absorbed by the ISM surrounding the O stars (i.e. case-B recombination). We compare the ionising rates for each \HII\ region to \citet[][Table 1]{Martins:2005bn} to estimate the spectral type and mass of a single MS star capable of producing each region, and report that in Table \ref{tab:Halpha_data}. \HII-4 and \HII-5 produce more ionising flux than a single O3 star, and thus are likely composed of multiple O stars (as noted, \HII-4 is made of three partially overlapping \HII\ regions). For example, the ionising flux from \HII-4 is equivalent to three O5V stars (each having $\sim \rm 37\, M_{\odot}$) and one O8V star ($\sim \rm 22\, M_{\odot}$), and \HII-5 has the equivalent ionising flux of two O5V stars and one O6V star ($\sim \rm 34\, M_{\odot}$). Other combinations of stellar masses and spectral types could equally well comprise the ionising output. In all cases, only a few O stars are needed to ionise each \HII\ region. We use the observed \HII\ regions and \Ha\ emission to determine the SFR using the calibration of \citet{meurer+09} adjusted to a Kroupa IMF. The SFR for each \HII\ region is listed in Table \ref{tab:Halpha_data}.

\subsection{Comparison with GALEX data}
\label{sec:galexcomp}

\begin{figure}
\centering
\includegraphics[width=85mm]{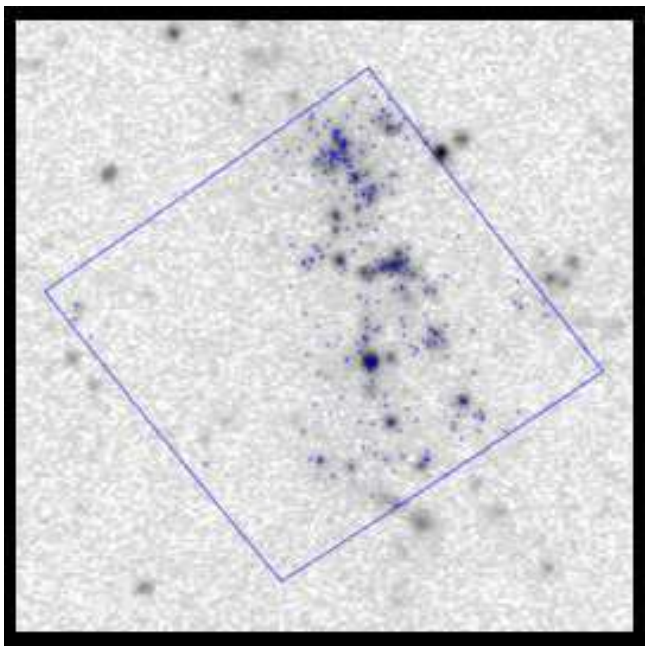} 
\caption[Comparison between GALEX FUV emission and MS stars.]{Comparison between the detected MS star positions from \textit{HST} imaging (blue dots) and FUV emission detected by GALEX FUV imaging (grey scale). The W3 field footprint is outlined. The FUV emission is well matched with the position of MS stars.
}
\label{fig:GALEX_comp}
\end{figure}

We have shown that the MS stars seen in the CMDs likely have masses $\ge 4 M_\odot$. These are B and O stars and should contribute strongly to the FUV emission of the outer disk as seen by GALEX. In Figure \ref{fig:GALEX_comp} we show the GALEX FUV image of M83's outer disk centred on W3 with the MS stars noted in Section~\ref{sec:CMDs} indicated. There is good agreement between the emission seen by GALEX and the MS stars.  However, some MS stars are too faint to be detected in this GALEX image which has a detection limit corresponding to the FUV emission from single B0 stars ($M_{\star} \approx 19\, \rm M_{\odot}$) at the $3\, \sigma$ detection limit  \citep{koda+12}. While lower mass stars contribute to the total UV light, they can not be individually detected at the depth of these GALEX images.

\section{Constraints on the initial mass function}
\label{sec:IMF_con}
Optical CMDs of young stellar populations generated from observations in two bandpasses are degenerate \citep[e.g.][]{Elmegreen:2006ho}; there is insufficient information to uniquely solve for both the IMF and Star Formation History (SFH).  Here we adopt the method developed by B15 of assuming a plausible SFH and using the MS luminosity function to constrain the IMF. The form of the IMF used throughout this paper is:
\begin{equation}
\xi(m) = \dfrac{dN}{dm} \propto m^{\alpha} \quad \text{for} \quad  1\, M_\odot < m <M_{u},
\end{equation} where $m$ is stellar mass in units of Solar masses. Hence, this is a power law IMF in which the lower mass limit is fixed at 1 $M_\odot$ (while lower mass stars presumably formed, we can not detect them individually, so they are ignored), and in which the upper mass limit, $M_u$, and power law index $\alpha$ are allowed to vary. In this form, the \citet{kroupa01} IMF has $\alpha=-2.35$ and $\rm M_u=120\, M_\odot$. 

\subsection{Choice of star formation history}
\label{sec:choiceSFH}

M83 has been well known for vigorous star formation in its optically bright portion for almost a century. Its peculiar bright core \citep{sp65,pastoriza75}, bares all the hallmarks of  
a starburst including wide spread intense narrow emission lines in the optical to infrared  \citep{bcz81, comte81, rouan+96, ccgk99}, intense continuum emission spanning the electromagnetic spectrum \citep[e.g.][]{tjd79, tfp85, skg87, bchs90, grltv91, rouan+96, ecw98, sw02, Buat:2002cg, Vogler:2005hy} with a spectrum dominated by young stellar populations \citep{bchss83, heckman+01, wlc11}, large quantities of dense molecular gas \citep{wbmp93, mhws99, dntww01, lwor04a, muraoka07} and numerous young blue star clusters \citep{bchs90, larsen99, lr00, Harris:2001bm, chandar+10, Andrews:2014co}, some with estimated masses up to those seen in Galactic globular clusters \citep{lwor04a}. Some areas in M83's central region are highly dust obscured, perhaps even the true nucleus (\citealp{diaz+06, rddad09}; however cf.\ \citealp{knapen+10}).  M83 also has a high observed supernova rate, having hosted six historical supernovae within the last century \citep[they are 1923A, 1945B, 1950B, 1957D, 1968L, 1983N;][]{botticella+12}, and containing numerous supernova remnants \citep{bl04, blair+14}. 

While the optically prominent disk is vigorously forming stars, the XUV disk outskirts appear to be forming stars more sedately (and we presume steadily) for at least 200 Myr \citep{thilker+05, thilker+07, Davidge:2010ib, gkr10}. 

On the basis of causality, the likely duration of a star formation event is related to its size.  That is, for short lived phenomena originating from the same event, the maximum extent of the phenomenon is related to its duration by the speed at which the event can spread. On the scale of the whole galaxy, M83's XUV disk displays arms to the North and South each extending beyond 30 kpc projected distance \citep{bigiel+10a}, suggesting SF lasting on time scale of the orbital time.  With an inclination corrected orbital velocity of about $160\, {\rm km\, s^{-1}}$ \citep{heald+16} and the radial position of the fields listed in Table \ref{tab:field_info}, the orbital times at their positions range from 530 to 950 Myr.  On a somewhat smaller scale, the clumpy distribution of MS stars spans structures that are still large compared to the $\sim 4.5$ kpc extent of each of the fields. Fig.~\ref{fig:M83_spat_dist} shows that the UV sources and MS stars are arranged in structures (typically arms) that traverse each of the fields. The \HI\ velocity dispersion in the outer disk is 12 to 18 km s$^{-1}$ \citep[][although these values may be somewhat inflated by beam-smearing]{heald+16}. A disturbance triggering star formation travelling at this speed would take 250 to 370 Myr to cross each field. In comparison, the MS stars that we are sensitive to have much shorter lifetimes $\le 150$ Myr.  Hence, the star formation within these fields likely lasted well over 100 Myr, for us to see it simultaneously across the kpc scales of each field and in fields separated by tens of kpc.  A roughly constant SFR over timescales of hundreds of Myr is thus a reasonable expectation of the true star formation history of the outer disk.

Long duration star formation is also consistent with the metal abundances in this portion of M83's XUV disk which can be produced with constant low-level star formation for $\approx$ 1-3 Gyr \citep{Bresolin:2009im,GildePaz:2007ij}. \citet{Bush:2008kg} showed that XUV disks can be successfully reproduced using simple prescriptions of star formation applied to an extended gas disk for a few Gyr.  This low-level continuous star formation would leave behind a disk of evolved stars, perhaps contributing to the populations of RGB stars seen in our fields (see Section \ref{sec:CMDs}). Hence the existing observations are all consistent with star formation lasting on the Gyr time scale.  This also corresponds to the orbital timescale of the extreme outer disks of galaxies \citep{mowzah18}.

As noted in B15, it is not just the temporal evolution of the SFR which is important for understanding upper-end IMF variations, but also the small-scale environment. It is a common assumption that all stars form in star clusters \citep[e.g.][]{ll03}. This assumption, however, may not hold in low density environments. It has been shown that protostars form over a large range of gas column densities and that there is no distinct break between clusters and the field population \citep{Gutermuth:2011he}. Low pressure environments should preferentially form stars in loose OB associations, not bound clusters \citep{elmegreen08}. The fraction of stars that form in bound star clusters has been shown to correlate with gas surface density and can be as low as 1\%\ in low surface-density environments, such as the outer disks of galaxies \citep{Kruijssen:2012bs}.  There is also strong evidence supporting the formation of isolated O stars in the Large Magellanic Cloud \citep{Bressert:2012gt}. Even in starburst galaxies, the UV light is dominated by diffusely distributed stars rather than compact star clusters \citep{meurer+95}.  This evidence suggests that not all stars form in clusters, and that the formation of bound star clusters is more rare in low density regions such as outer disks. The distribution we find of MS stars (Fig.~\ref{fig:M83_spat_dist}), while clumpy, is spread across the 4.6 kpc width of each of our fields, i.e.\ much bigger than a single star cluster (effective radius $\lesssim 1$ pc, tidal radius $\lesssim 10$ pc)\footnote{We reiterate that our visual inspection of our HST images did not reveal any compact clusters missed by our source finding. Our experience is that at the distance of M83, and depth of our HST observations, compact clusters either are typically partially resolved in to individual stars at their outskirts \citep[e.g.\ the globular clusters in NGC~2915][]{meurer+03} or are measurably less concentrated than unresolved sources \citep[presumably isolated stars, binaries, or small multiple star systems, e.g.][]{cook+19}.}.  We also combine the data from our four fields in the subsequent IMF analysis to smooth over any local enhancements that may occur in the individual fields.  Thus, following B15, we model star formation in the outer disk as non-clustered, random sampling of the IMF and SFH.  In Section~\ref{sec:non_const_SFH} we relax this assumption and test whether combined burst and continuous star formation models can account for the observed MSLF.

\subsection{Simulated colour-magnitude diagrams}
\label{sec:sim_CMDs}
Various groups have used HST images of resolved stellar populations to constrain properties of the populations, by simulating the entire CMD \citep[e.g.][and references therein]{dalcanton+09,weisz+11a,tolstoy+09}.  Here, as in B15, we concentrate on the MS, as the best understood phase of stellar evolution. Later stages are highly sensitive to stellar rotation \citep{vlsmm07,ekstrom+12,georgy+12,georgy+13,Leitherer:2014tv,lemsal14}, metallicity \citep{eit08,wte15,gdmg17}, and binary star evolution effects \citep{eit08,gdmg17}. We use the MSLF to constrain the upper-end of the IMF. Here we employ ``traditional'' evolutionary tracks of non-rotating, windless single stars. We use the the $I$ versus $B-I$ CMD as this has the largest colour baseline from our data, making the MS stars easier to separate from other evolutionary phases. To determine the best-fitting upper-end IMF in the outer disk we first produce an ensemble of simulated CMDs in which we vary the IMF slope ($-3.95 \leq \alpha \leq -1.95$), and upper-mass limit ($\rm M_u  = $ 15, 20, 25, 40, 60, 85, 120 $M_\odot$). 

A brief outline of the method used to model the CMDs is given below, details can be found in B15 (their sections 4.2 and 4.3). To produce the simulated CMDs we randomly sample stars from the assumed IMF and adopt a constant SFR over 300 Myr.  This duration is chosen to be twice the MS lifetime of the lowest mass stars in the MS selection polygon so as to allow low mass stars to scatter into the selection box after applying simulated errors. To model stellar evolution we use the PARSEC evolutionary tracks \citep{Tang:2014gt, Bressan:2012bx}, which again assume single, non-rotating, windless stars.  While the stellar evolution phenomena noted above should also affect the MS, we do not expect the effects to be as severe as for the later evolutionary stages.  This approach also allows a more direct comparison to the previous resolved stellar population studies. We assume uniform foreground Milky Way dust extinction with reddening $\rm E(B-V)=0.06$, from \citet{Schlafly:2011iu} and adopt the \citet{Fitzpatrick:1999dx} reddening law with $\rm R_V = 3.1$, and uniform metallicity $\rm Z\sim 0.3\,Z_\odot$ \citep{Bresolin:2009im}. The foreground extinction is comparable to the average total extinction found in the outer disk by \citet{Bresolin:2009im} ($\rm E(B-V)=0.05$ for $\rm r>r_{25}$), hence we do not include any internal dust correction. Except for the constant SFR, these are common assumptions of CMD analyses in nearby galaxies \citep[e.g.][]{williams+08, annibali+13, lc13, meschin+14}. In Section \ref{sec:caveats}, we discuss the biases these assumptions may induce in our results.

We use the PARSEC stellar evolutionary tracks, which have been extended to high-mass stars for low metallicities \citep{Tang:2014gt, Bressan:2012bx} and interpolate between the available tracks at the same evolutionary phase to determine the surface gravity and effective temperature at the required age for each star. To determine the observed magnitudes in each filter, the surface gravity and effective temperature is matched to a grid of stellar atmospheres. We use grids of stellar-model colours, which are corrected for extinction and constructed in the HST filters (Bianchi, in preparation) tabulated in part by \citet{Bianchi:2014fk} and shown by \citet{Bianchi:2008dq}. We then interpolate between grid points to determine the magnitude in each filter. Approximately 0.1\%\ of the simulated stars have higher temperatures than our stellar atmosphere grid; for those we extrapolate the available data. We model photometric errors and correct for completeness using the results from the artificial star tests.  That is, we perturb the modelled photometry by the median expected error times a random variable with Gaussian distribution having a mean of zero and a dispersion of one, and we randomly remove a fraction of the stars to match our completeness tests.  This is done using the field specific noise parameters determined from the artificial star tests. As demonstrated by Figures~\ref{fig:M83_AST_err} and \ref{fig:M83_CMDs} the errors are manageable and the completeness corrections small over the range of stellar brightnesses most relevant to this study, that is covering the MS, as well as the BHeB sequence.  

At the end of each simulation we extract the MS stars from the $I$ versus $B-I$ CMD using the same MS selection box as the observations.  The MSLF is the distribution of the \textit{I\/} band magnitudes of the MS stars in the selection box. We use the \textit{I\/} band MSLF because it provides smaller errors at a given apparent magnitude and is less affected by dust extinction than the other bands. We produce four batches (one for each field, each initiated with a different random number seed) of 100 simulations, for each set of IMF parameters to account for the effects of stochasticity. The number of MS stars in each simulation is matched to that observed in the field it is to be compared to. We compare the simulated MSLFs to the observed MSLF, after casting them in to cumulative form, using the Kolmogorov--Smirnov (K--S) test. We use the K--S test statistic $d$, averaged over the simulations of each set of IMF parameters, as a measure of how well the simulated MSLF matches the observations. Our assumption is that IMF parameters producing the minimum average $d$ correspond to the IMF closest to that in our observations. As usual, for the given number of data points and $d$, we calculate the probability $p$ that the simulated and observed distribution were randomly drawn from the same parent distribution. We perform repeatability tests using the simulated data in order to constrain the errors, as detailed in Section \ref{sec:errs}.

\subsection{IMF constraints from the MSLF}
\label{sec:IMF_constraints}

\begin{figure}
\centering
\includegraphics[scale=0.45]{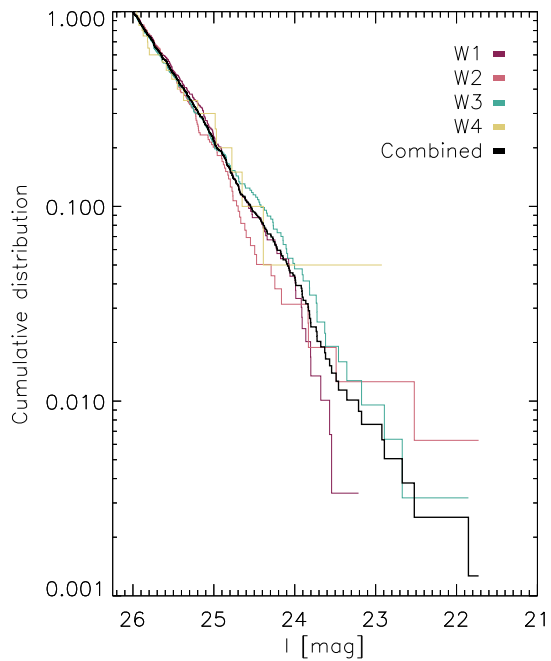}
\caption{Comparison between the observed MSLF for each of the fields and the combined MSLF. Each MSLF is shown as the cumulative distribution of MS stars as a function of \textit{I}\/ band magnitude. \label{fig:M83_IMF_compare_comb}}
\end{figure}

To test if there are significant differences between the MSLFs in the four fields we used K--S tests for each pair of fields. The value of $p$ ranges from 0.36 (W1 compared to W2) to 0.74 (W2 compared to W4), hence we do not detect any significant differences in the MSLFs between the fields. Therefore, we combine the observed MS stars from each of the four fields to produce a combined MSLF.  This allows us to better reproduce our assumption of a constant SFH; by combining the fields we average over any local variations in the SFR. This also increases the number of stars compared to the single field analysis, improving the accuracy of the constraints on the IMF parameters, especially $\alpha$ (B15). Figure \ref{fig:M83_IMF_compare_comb} shows the combined observed MSLF along with the individual MSLFs for each field. To do the comparison with simulated data we combine the batches of simulations for each field. 

\begin{table}
\centering
\caption{K--S test results comparing the observed combined MSLF to the best-matching realisation of a simulation with Kroupa IMF and the best-fitting IMF ($\alpha=-2.35 $ and $\rm M_{u}=25\,M_\odot$). In the case where we compare simulations to observations we list the median value from the 100 realisations. We consider the threshold for a significant difference between MSLFs to be $p < 0.01$. \label{tab:MSLF_compare}}
\begin{tabular}{c|c|c|c}
\hline
MSLF 1     & MSLF 2           & $d$   & $p$   \\
\hline \hline
Combined   & best-fitting MSLF  & 0.037 & 0.65  \\    
Combined   & best-fitting $w_{\rm H\alpha}$ & 0.057 & 0.15 \\ 
Combined   & Kroupa             & 0.086 & 0.006 \\ 
\hline
\end{tabular}
\end{table}

\begin{figure}
\includegraphics[scale=0.4]{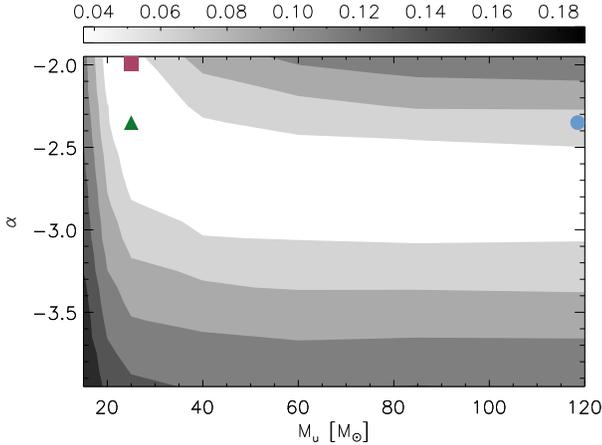}
\caption{Contour plot showing the mean test statistic, $d$, for the IMF slope $\alpha$ and upper-mass limit $\rm M_{u}$ for the combined MSLF IMF analysis. The best-fitting IMF parameters (where $d$ is minimised) are shown as a filled olive-green triangle.  There is however, little difference between the best-fitting parameters and the elongated minimum region shown in white. The Kroupa IMF parameters are indicated with a pale-blue filled circle, while the best-fitting IMF parameters that match the \Ha\ constraints are indicated with a brick-red filled square.  These latter two symbols have been shifted slightly from their nominal positions at the edge of the parameter space so as to be clearly visible in the figure.\label{fig:M83_comb_contour}}
\end{figure}
    
Figure \ref{fig:M83_comb_contour} shows the results of the K--S test as a contour plot of $d$ in the plane of the free parameters, the IMF slope ($\alpha$), and upper-mass limit ($\rm M_{u}$). The lower the value of $d$ the better the fit, with the parameter set with the lowest $d$ corresponding to the best fit. The best-fitting parameters are $\alpha=-2.35 $ and $\rm M_{u}=25\,M_\odot$ (hereafter the best-fitting IMF parameters). This figure also reveals an extended minimum region which encompass $40 <\rm M_{u}/M_\odot<120$ with $\alpha \sim -2.85$, indicating that the upper-mass limit is not well constrained. High-mass MS stars have a significant change in luminosity over their MS lifetime with limited change in optical colour, i.e.\ their evolutionary tracks are steep, making it hard to differentiate stellar masses using optical photometry alone. 

\begin{figure}
\centering
\includegraphics[scale=0.45]{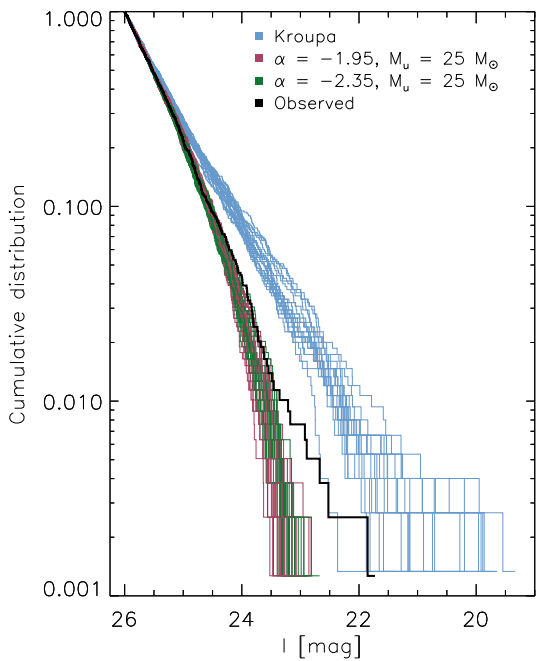} 
\caption{Comparison between the combined observed MSLF (black) and 20 of the closest matching realisations for a Kroupa IMF (pale-blue), best-fitting IMF parameters from the MSLF analysis (olive-green; Section \ref{sec:IMF_constraints}), and best-fitting IMF parameters determined via matching the observed \Ha\ equivalent width to simulations (brick-red; Section \ref{sec:IMF_constraints_Ha}). Each combined MSLF is shown as the cumulative distribution of MS stars as a function of $I$ band magnitude. As detailed in Section~\ref{sec:sim_CMDs}, the simulations differ in which stellar masses are randomly selected, the random addition of photometric errors and the random removal of stars to match the artificial star test results detailed in Section~\ref{sec:artstar}.  The differences at the high-luminosity end of the MSLFs appear exaggerated by the logarithmic scale used. \label{fig:M83_IMF_compare}}
\end{figure}

Figure \ref{fig:M83_comb_contour} reveals that an IMF deficient in high-mass stars compared to the Kroupa IMF is preferred to fit the MSLF in the outer disk of M83. In Figure \ref{fig:M83_IMF_compare} we compare the observed and simulated MSLFs, plotted as normalised cumulative distributions, for the 20 best-matching realisation to the observed data for each of three sets of IMF parameters: (1) the Kroupa IMF; (2) $\alpha=-2.35$ and $\rm M_u=25 \, M_\odot$; and (3) $\alpha=-1.95$ and $\rm M_u=25 \, M_\odot$. Set (2) corresponds to the best-fitting IMF parameters from the MSLF analysis (Figure \ref{fig:M83_comb_contour}), while set (3) are the best-fitting IMF parameters from the \Ha\ analysis in Section \ref{sec:IMF_constraints_Ha}, below (Figure \ref{fig:M83_Ha_compare}).  The best-fitting IMF is clearly preferable to the Kroupa IMF which we confirm using the K--S test. We list the K--S test statistic $d$, and $p$ for the best-matching realisations for the Kroupa and our preferred IMF in Table \ref{tab:MSLF_compare}. In this case the K--S test rules out a Kroupa IMF ($\rm p=0.006$) and we show that the best-fitting IMF is well matched to the combined observed MSLF. 

\subsection{Uncertainties in the best fit IMF parameters}
\label{sec:errs}

While Fig. \ref{fig:M83_comb_contour} illustrates the range of plausible fits with similar $d$ values in the K--S test, it does not indicate the uncertainty in the IMF parameters. To address this, we use simulations to determine how well our technique recovers known IMF parameters, and to estimate their uncertainties.  We do this by selecting a simulated stellar population (as described in Section~\ref{sec:sim_CMDs}) as the `observed' data (with known $\alpha$ and $\rm M_u$). The `observed' IMF is then randomly selected MS stars matched in number to the real observations.  We then use the same K--S test minimisation to compare this MSLF with all the remaining simulations at this particular set of $\alpha$ and $\rm M_u$, as well as the simulations at each grid point. We repeat this for the other 99 simulations with the same $\alpha$ and $\rm M_u$ as the `observed' MSLF. The distribution of the recovered values gives the uncertainty at that particular set of IMF parameters.  As is often the case when fitting an arbitrary function, the distribution of recovered values is not  Gaussian in detail.  This is especially so since the simulations are made at fixed grid points, hence the recovered values are strongly discretised.

Using the one hundred MSLF simulations with our best fit parameters as the input to this process yields the most common recovered IMF slope via the K--S test to be $\alpha = -2.35 \pm 0.3$, where the uncertainty encompasses the 16th to 84th percentile of the distribution of recovered values.  The same simulations recover the best $M_u = 25\, M_\odot$ in 72 of the simulations including all of those in the 16th to 84th percentile range.  Thus our grid of simulations is too coarse to properly determine the 1 $\sigma$ uncertainty in $M_u$. In a normal distribution, the 2nd and 98th percentiles of the recovered values provide the 2 $\sigma$ confidence interval.  For our simulations these percentiles in $M_u$ are 20 $M_\odot$ and 60 $M_\odot$.  Scaling the difference between these and the best fit values yields our adopted result and estimated 1$\sigma$ errors as $M_u = 25^{+17}_{-3}\, M_\odot$. 

\subsection{IMF constraints from the \Ha\ observations}
\label{sec:IMF_constraints_Ha}

As done in B15, we use \Ha\ fluxes (Section \ref{sec:HII_regs}) as an additional constraint on the IMF.  We assume the \HII\ regions are undergoing case B recombination (i.e.\ they do not ``leak'' ionising photons) and that our apertures encompass all the \Ha\ emission, and hence that the \Ha\ flux of the \HII\ regions gives an estimate of the total ionising flux of the fields.  If ionising photons escape from the galaxy, or even just beyond the \HII\ regions, than we will underestimate the O star content.  The possibility of ionising photon leakage is discussed further in Section~\ref{sec:caveats}.  In order to compare the \Ha\ flux with the stellar population most tied to its ionisation, and to minimise the effects of distance uncertainties, we form a `pseudo' \Ha\ equivalent width, $w_{\rm{H} \alpha}$ by dividing the \Ha\ flux by the summed $V$ band flux densities of the MS stars identified in Section~\ref{sec:CMDs}.  This is not a true equivalent width because the continuum flux density is not at the same wavelength as \Ha, not all the stellar populations within the fields are included, and because the MS star selection has an arbitrary lower luminosity limit.  Since we are concerned with just the young stellar populations and hot MS stars have a fairly flat optical spectrum, incompleteness is our major concern.  This means that $w_{\rm{H} \alpha}$ will overestimate the true \Ha\ equivalent width of the young stellar populations.  To account for this incompleteness, we use the same MS selection box in our simulations as in our observations. Table \ref{tab:EW} lists $w_{\rm{H} \alpha}$ for each of the fields and integrated over all fields. Table~\ref{tab:EW} compiles these along with other measurements related to the integrated star formation rate of the fields.  These include the number of MS stars, the absolute magnitude in the FUV ($M_{\rm FUV}$), and the (logarithm of) the \Ha\ luminosity ($L_{\rm H\alpha}$).  

\begin{table}
\centering
\caption[Integrated star formation rate indicators for each field in M83]{Integrated star formation rate indicators for each field in M83. \label{tab:EW}}
\begin{tabular}{|c|r|c|c|c}
\hline
Field & $N_{\rm MS}$ & $M_{\rm FUV}$ & $\log(L_{\rm H\alpha})$ & $w_{\rm{H} \alpha}$ \\
~     & ~          & (ABmag)    & (log(erg s$^{-1}$)) & (\AA) \\
\hline \hline
W1 & 296 & $-11.53 \pm 0.07$     & $37.96 \pm 0.01$ & $396 \pm 7$   \\
W2 & 159 & $-10.96 \pm 0.12$     & $37.92 \pm 0.01$ & $647 \pm 7$  \\
W3 & 314 & $-11.61 \pm 0.07$     & $37.88 \pm 0.03$ & $298 \pm 9 $  \\
W4 &  20 & $-9.04^{+1.18}_{-0.55}$ & $\ldots$ & 0 \\
\hline
\end{tabular}
\end{table}

We employ the same simulations described in Section~\ref{sec:sim_CMDs} to model $w_{\rm H\alpha}$.  For each simulation, we randomly select stars meeting our MS selection criteria, match the observed number of MS stars for each field, and combine the results for all fields.  As in B15, we use Table 3.1 from \citet{Conti:2008ur} to estimate the ionising output of each MS star according to its initial mass, and then convert these to the equivalent \Ha\ luminosity.  These are summed, as are the corresponding modelled $V$-band luminosities, and the ratio of the two taken to form the modelled $w_{\rm H\alpha}$ for the simulation. 

We count the number of simulations at each set of IMF parameters that reproduce the observed $w_{\rm H\alpha}$ within 20\%\ as those that ``match''.  This 20\%\ criterion is somewhat arbitrary and chosen to be consistent with the work of B15.  There we note that employing a matching constraint based on observational errors may be too tight, while one based on the actual number of ionising stars may be preferable but is not known {\em a priori\/}. 

\begin{figure}
\centering
\includegraphics[scale=0.4]{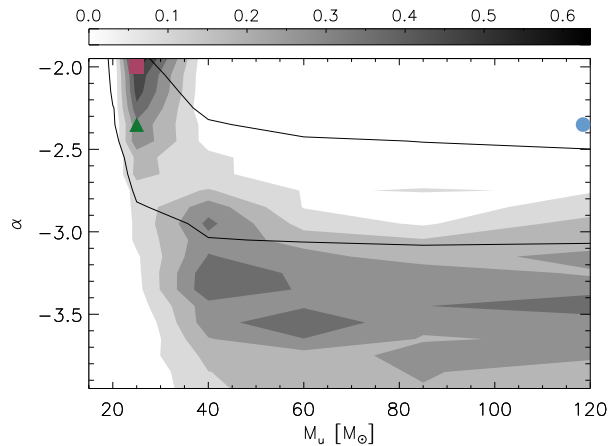} 
\caption{Contour plot showing the fraction of simulations which have a pseudo \Ha\ equivalent width ($w_{\rm H\alpha}$) matched (within 20\%) to the combined observed $w_{\rm H\alpha}$. The IMF parameters with the maximum number of matches are indicated with the brick-red filled square. We show the best-fitting IMF parameters determined via the MSLF analysis as the olive-green filled triangle, and the Kroupa IMF parameters with a pale-blue filled circle. Note, the square and circle symbols have been shifted slightly from their nominal positions at the edge of the parameter space so as to be clearly visible in the figure.  The best-fitting region from Figure \ref{fig:M83_IMF_compare} is shown by the black contour for comparison. \label{fig:M83_Ha_compare}}
\end{figure}

In Figure \ref{fig:M83_Ha_compare} we show the fraction of simulations which have a $w_{\rm H\alpha}$ that we consider a match to the observed $w_{\rm H\alpha}$. To ease comparison to the MSLF analysis we show the best-fitting region from the MSLF fitting (Fig.\ \ref{fig:M83_comb_contour}) as a black contour. The IMF parameters with the highest percentage of matches (74\%) are $\alpha=-1.95$ and $\rm M_u=25 \, M_\odot$. The figure indicates that the simulations using the best fit IMF parameters from the MSLF also have a high rate of matching (36\%) the $w_{\rm H\alpha}$ observations. There is a second region of enhanced $w_{\rm H\alpha}$ matching corresponding to $\alpha \sim -3$ to --3.5, which is ruled out by the MSLF analysis. 

\begin{figure}
\centering
\includegraphics[scale=0.45]{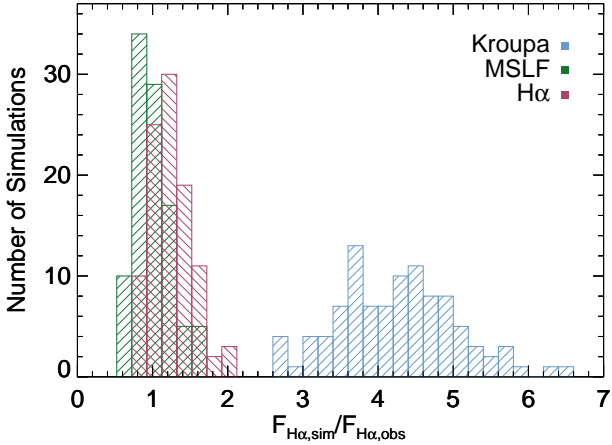}
\caption[Distribution of the ratio of simulated \Ha\ flux to  observed \Ha\ flux]{Distribution of the ratio of simulated \Ha\ flux ($\rm F_{H\alpha, sim}$) assuming a Kroupa IMF to the observed \Ha\ flux for 100 simulations for each of three different IMF parameter sets, as indicated by the legend.} \label{fig:Rat_Ha}
\end{figure}

Figure \ref{fig:Rat_Ha} shows the distribution of the ratio of the simulated \Ha\ flux to that observed, for each of three sets of simulations.  Here, the total observed \Ha\ flux over the four fields is $\rm F_{H\alpha,obs}= 8.95 \pm 0.11 \times 10^{-14} \rm erg\, cm^2\, s^{-1}$ (see Table \ref{tab:Halpha_data}).  The IMF parameters for the three sets of simulations shown are the standard Kroupa values, those that best fit the MSLF, and those best-matching the $w_{\rm H\alpha}$ observations (shown as the pale-blue circle, olive-green triangle, and brick-red square, respectively, in Fig.~\ref{fig:M83_Ha_compare}). For a Kroupa IMF, on average, the expected \Ha\ flux is $4.2 \pm 0.8$ times larger than the observed \Ha\ flux across the observed fields, with a minimum ratio of 2.6. Hence, simulations with a Kroupa IMF consistently over predict the observed \Ha\ emission.  This is because they produce too many of the highest mass ionising stars.  In contrast, the latter two sets of IMF parameters yield  $F_{\rm H\alpha,sim}/F_{\rm H\alpha,obs}$ ratios closest to unity; either set of parameters produce \Ha\ fluxes largely consistent with that observed.  This result is expected for the IMF parameters matching the $w_{\rm H\alpha}$ observations.  However, the MSLF fitting is not constrained by the \Ha\ data, yet produces a population of high-mass stars that well match the \Ha\ observations. Thus, the paucity of \Ha\ emission in M83's XUV disk is consistent with the MS stars seen by HST. 

It is interesting to compare our results to those of \citet{koda+12}, who find that a stochastically sampled Kroupa IMF and an ageing effect can explain the \Ha/FUV flux ratio observed in star clusters within the XUV disk of M83. While \citet{koda+12} study the light only from cluster populations, we observe and include young stars with $\rm M_{\star}>4\, M_\odot$ (i.e.\ O or B stars) both in `clusters\footnote{We use the term cluster loosely to mean stars that appear to be spatially grouped.}' and spread diffusely over the field. The FUV light from the diffuse stellar populations is likely missed by Koda et al.\ who only consider sources found using \textsc{SExtractor} on the GALEX FUV image with a detection limit corresponding to a single B0 star, or equivalently a typical young star cluster with a mass of at least 100 M$_\odot$.   The majority of the B star sequence will be missed by Koda et al.\ if they are diffusely spread or in smaller mass clusters. Figure~\ref{fig:M83_spat_dist} shows the spatial distribution of the various types of stars identified with our CMD analysis as well as the clusters indicated by \citet{koda+12}.  While there are MS stars at the positions of almost all their clusters, the majority of MS stars are scattered well beyond them.  Similarly, the GALEX images shows considerable diffuse UV emission beyond the bright peaks \citep[i.e.\ the clusters identified by][]{koda+12}.  This is shown in Fig~\ref{fig:GALEX_comp} where it is evident that this diffuse UV glow encompasses almost all the MS stars in our HST images. This neglected UV light may considerably reduce the \Ha/FUV ratio integrated over the entire XUV disk compared to that found in the UV brightest cluster, and thus increase the difference between the models and observations of \citet{koda+12}.

\subsection{Comparison of Star Formation Intensity Estimates}

Table~\ref{tab:sfcomp} compiles field by field estimates of the face-on star formation intensity (expressed in the log) for the three sets of upper end IMF parameters highlighted in this study: the Kroupa IMF; our best fit to the MSLF; and our best match to $w_{\rm{H} \alpha}$; and the three sets of high-mass star formation rate indicators listed in Table~\ref{tab:EW}: the number of MS stars ($N_{\rm MS}$); the far ultraviolet absolute magnitude $M_{\rm FUV}$; and the \Ha\ luminosity ($L_{\rm H\alpha}$). As done throughout this paper, here we assume $m_l = 1$ M$_\odot$.  To convert these quantities into logarithmic SFR, one should add 1.35 dex, corresponding to the area in kpc$^{2}$ in the disk plane of the WFC observations.  The scaling between $N_{\rm MS}$ and SFR is based on the CMD simulations made for this study. The scaling between SFR and luminosity in the FUV and \Ha\ is based on Starburst99 spectral synthesis modelling \citep{leitherer+99,vl05,leitherer+10,lemsal14}.  The modelling used the Padova group evolutionary tracks and standard mass loss rates at LMC metallicity \citep[$Z = 0.008$ close to that of the outer disk of M83; Table~\ref{tab:M83_prop};][]{mmssc94,smms93}. Note these evolutionary tracks were created by the same research group that produced the PARSEC isochrones used to model the CMDs, albeit the tracks used in Starburst99 predate those used in the CMD analysis by about 20 years. Experiments with models using the (newer) Geneva group evolutionary tracks at LMC metallicity produced scalings consistent to those used in Table~\ref{tab:sfcomp} within 0.03 dex.  However, solar metallicity models yield a higher SFR by 0.1 to 0.2 dex compared to LMC metallicity models when the Padova tracks are employed.  

In general, the star formation intensity estimated from $N_{\rm MS}$, and $M_{\rm FUV}$ track each other well (to about 0.2 dex), while the \Ha\ based star formation intensity is more discrepant (lower), particularly when the Kroupa IMF is adopted.  This is sensible, in that the GALEX fluxes are dominated by the MS stars, while the \Ha\ fluxes are the summed fluxes from \HII\ regions which neglects any faint \Ha\ emission there may be beyond the boundaries we have drawn.  A comparison of column 13 in Table~\ref{tab:Halpha_data} and the second column of Table~\ref{tab:EW} shows that the fraction of MS stars that fall within the boundaries of \HII\ regions ranges between 36\%\ and 53\%\ for fields W1, W2, and W3 (47\%\ averaged over the three fields), while W4 has no \HII\ regions, and also very few MS stars.  This spatial mismatching of the the recently formed stars and the \HII\ regions explains, in part, the discrepancy of the \Ha\ based estimate of star formation intensity and that from the other two tracers. The poor matching of the different star formation intensity estimates when adopting the Kroupa IMF is consistent with the poor fits to the MSLF and $w_{\rm{H} \alpha}$ for this IMF.  

The star formation intensities listed in Table~\ref{tab:sfcomp} are remarkably low compared to what is typically found in star forming galaxies, no matter which IMF is adopted.  In order to ease comparisons with previous studies, in this paragraph we adopt the Kroupa IMF results.  For fields W1, W2, and W3, the star formation intensity, estimated by MS stars or the UV flux, of $\sim 10^{-4.2}$ is 2.5 dex weaker than typically found in disk galaxies \citep[as calculated from the \Ha\ effective surface brightness corresponding to the median contribution to the volume averaged \Ha\ emissivity of the local Universe as sampled by the SINGG survey][]{hanish+06,audcentross+18}. Table~\ref{tab:sfcomp} shows the star formation intensity in the W4 field is about a factor of 10 times even more dilute than the other fields, closer to the radially averaged star formation intensity of the outer disk.  The contrast with more intense star forming environments is more extreme. Compared to the median (50th percentile) effective surface brightness of starburst galaxies in the nearby Universe, as observed in the dust corrected UV \citep{meurer+97}, the star formation intensity of fields W1 to W3 is 5 dex less intense.  These fields are a further 0.8 dex fainter than the 90th percentile UV surface brightness of starbursts - the ``Starburst Intensity Limit'' of \citet{meurer+97}.  Starbursts contain numerous star clusters, which provide a significant fraction ($\sim20$\%) of their UV flux \citep{meurer+95}, but are much smaller (effective radii on the order of 1 pc or smaller).  Their effective UV surface brightnesses are more than 1200 times more intense than the starbursts that contain them, hence they are over 8 dex more intense than these outer disk fields.  This is a lower limit because the median size of the clusters in \citet{meurer+95} has not been measured.

\begin{table*}
\centering
\caption{Estimates of the log of the face-on star formation intensity for the four fields in units of $M_\odot$ kpc$^{-2}$ year$^{-1}$.  Three sets of upper end IMF parameters are considered: Kroupa, MSLF best fit, and \Ha\ best fit (the relevant upper end IMF parameters are listed parenthetically on the second header line of the table). For each set of IMF parameters, the star formation intensities are listed as estimated from the number of main-sequence stars ($N_{\rm MS}$), the absolute magnitude in the far ultraviolet ($M_{\rm FUV}$), and the \Ha\ luminosity ($L_{\rm H\alpha}$). \label{tab:sfcomp}}
\begin{tabular}{l|ccc|ccc|ccc}
\hline
Field & \multicolumn{3}{c}{Kroupa} & \multicolumn{3}{c}{MSLF best fit} & \multicolumn{3}{c}{\Ha\ best fit} \\
~ & \multicolumn{3}{c}{($M_u=120\, M_\odot$, $\alpha=-2.35$)} &
\multicolumn{3}{c}{($M_u=25\, M_\odot$, $\alpha=-2.35$)} &
\multicolumn{3}{c}{($M_u=25\, M_\odot$, $\alpha=-1.95$)} \\
~ & $N_{\rm MS}$ & $M_{\rm FUV}$ & $L_{\rm H\alpha}$ & $N_{\rm MS}$ & $M_{\rm FUV}$ & $L_{\rm H\alpha}$ & $N_{\rm MS}$ & $M_{\rm FUV}$ & $L_{\rm H\alpha}$ \\ 
\hline \hline
W1       & $-4.16$ & $-4.14$ & $-4.67$  & $-4.21$ & $-4.13$ & $-3.76$  & $-4.38$ & $-4.26$  & $-4.02$  \\
W2       & $-4.43$ & $-4.37$ & $-4.70$  & $-4.47$ & $-4.35$ & $-3.80$  & $-4.64$ & $-4.49$  & $-4.05$  \\
W3       & $-4.13$ & $-4.11$ & $-4.74$  & $-4.18$ & $-4.10$ & $-3.84$  & $-4.35$ & $-4.23$  & $-4.09$  \\
W4       & $-5.34$ & $-5.14$ & $\ldots$ & $-5.38$ & $-5.12$ & $\ldots$ & $-5.55$ & $\ldots$ & $\ldots$ \\
\hline
\end{tabular}
\end{table*}

\subsection{Caveats and limitations}
\label{sec:caveats}
The results presented here depend on the assumptions of our model, as well as the corrections applied to our data.  

As listed in Table~\ref{tab:M83_prop}, we adopt the distance $D = 4.5$ Mpc following \citet{Karachentsev:2002fs}, based on HST observations of the tip of the Red Giant Branch. The NASA Extra-Galactic Database (NED) lists four estimates of $D$ based on HST observations on the tip of the RGB ranging from 4.51 to 4.92 Mpc \citep{Karachentsev:2002fs,jacobs+09,radsmi+11,tully+13}.  Estimates based on Cepheid variable stars have a narrower range of $D$ ranging from 4.50 to 4.66 Mpc \citep{thim+03,sttrs06,tully+13}.  Thus, our adopted $D$ is at the low end of modern estimates of the distance to M83, but nevertheless consistent with measurements of both Cepheid and tip of the RGB stars.  The spread in these estimates, 4.5 to 4.92 Mpc, amounts to 0.19 mag in luminosity, and can be directly mapped in to an uncertainty in $M_u$.  In comparison, for our adopted evolutionary tracks, and at equivalent phases of evolution during the main-sequence phase, stars with an initial mass of 25 $M_\odot$ will be 0.35 mag brighter in $B$, $V$, and $I$ than those with an initial mass of 20 $M _\odot$.  Hence, the uncertainty in $D$ is smaller than the separation between our fiducial models, and will not shift our $M_u$ estimates outside the quoted range of uncertainty $M_u = 25^{+17}_{-3}\, M_\odot$.

In our model we assume uniform dust extinction ($E(B-V)=0.06$ mag; Table~\ref{tab:M83_prop}), equivalent to just the foreground Galactic dust extinction. This results in the observed position of the main-sequence and BHeB sequence to be in good agreement with the colour-magnitude diagrams, as shown in Fig.~\ref{fig:cmd_all} and is consistent with the {\em average\/} total reddening of the \HII\ regions in our fields studied by \citep{Bresolin:2009im}.  In contrast, \citet{GildePaz:2007ij} list total reddening ranging from $E(B-V) = 0$ to 0.29 mag, with an average of 0.13 mag in four XUV regions in our W3 field.  The {\em maximum\/} dust reddening of these XUV regions corresponds to an additional reddening of $\Delta E(B-I) = 0.53$ mag, $\Delta E(V-I) = 0.22$ mag compared to our adopted reddening.  MS stars with this reddening would have colours redder than the observed BHeB sequence.  Since the ISM is likely to be particularly dense, clumpy, and dusty where new stars are being formed \citep[e.g.][]{Pellegrini:2012ii} this could contribute intrinsic scatter and a bias to the photometry, which we have not modelled. We note that an intrinsic scatter in dust reddening should manifest itself as a blurring between the MS and the BHeB sequence \citep{mcquinn+11,lc13} which is not apparent in our CMDs (Figs~\ref{fig:M83_CMDs} and \ref{fig:cmd_all}). This is consistent with \citet{dong+08} who find the outer disk of M83 (specifically in two fields containing our W2 and W3 fields) is weak in near to mid infrared emission, which suggests sight-lines to shrouded high-mass star formation are rare.  They note that if the assumed dust extinction of sources matched in both the UV and infrared is as high as the most reddened XUV sources found by \citet{GildePaz:2007ij} then they should be very young with a strong ionizing spectrum (for a presumed Salpeter like IMF) while the observed lack of significant \Ha\ emission would require a high escape fraction of ionizing photons.  This would require an ISM geometry or composition that allows dust to redden the XUV sources but not capture the higher energy ionizing photons.  We are not aware of such structures being shown to exist in astrophysics.

A similar concern regarding dust extinction involves the sources just to the red of our MS selection box: some of these are more luminous in the CMDs (figures~\ref{fig:M83_CMDs} and \ref{fig:cmd_all}) than the brightest MS stars.  We have presumed that these are BHeB stars, but perhaps these are the most-massive stars which are under-counted because they are preferentially slightly reddened out of our MS selection box.  If so, we would expect them to be more prevalent in the \HII\ regions.  However, limiting ourselves to those with $m_I \leq 23.75$, and covering the colour range $B-I = -0.27$ to 0.5 we find that 34\%\ of the brightest BHeB stars are within the boundaries of the \HII\ regions within the fields W1, W2, and W3, almost exactly equal to the 33\%\ of MS stars in those fields within the boundaries of the \HII\ regions.  This is consistent with them being stars that have relatively recently evolved off of the MS as we expect for BHeB stars with the same spatial distribution as the current MS stars.  Hence, we do not find evidence supporting the notion that the most luminous stars slightly to the red of our MS selection box are reddened MS stars.

As noted in Section~\ref{sec:obs_phot}, the four fields we observed were chosen in part by UV brightness.  This could bias our results to regions with recent star formation and hence high O star content.  However, this possible bias is countered by selecting fields to cover a range of UV surface brightnesses.  It may be that the latter selection criterion drives our results towards a deficiency in the most massive stars.  As noted in Section~\ref{sec:IMF_constraints} there are no significant differences in the MSLF between the fields, hence we deem it unlikely that field selection is driving our results.  Nevertheless, similar observations covering a larger continuous area of the outer disk of M83 would provide a more convincing demonstration that field selection is not driving our results.

As in our study of the outer disk of NGC~2915 (B15), we employ evolutionary tracks of non-rotating stars.  However, it has long been known that massive stars, especially B stars, typically are found to be strongly rotating (\citealt{morgan44,slettebak49}; for more recent studies see e.g.\ \citealt{hunter+08,hunter+08err,zr12,dufton+13}).  Rotation has a strong effect on the evolution of stars, and consequently on the CMDs of populations of massive stars, and hence on how the CMDs should be interpreted.  Most relevant to this study, rotating stars have a longer MS lifetime, are hotter, and more luminous than their non rotating counterparts \citep{ekstrom+12,Levesque:2012bv,Leitherer:2014tv,georgy+13}.  The effects increase with metallicity and are more pronounced in the post MS phases of evolution.  By neglecting rotation, our results will be biased towards higher $M_u$ and flatter $\alpha$.  Since the metallicity of the outer disk of M83 is low \citep[i.e.\ $\rm Z=0.3\, Z_\odot$][]{Bresolin:2009im}, the effect of stellar rotation should be reduced compared to solar metallicity models.

Similarly, we note that our results have a sensitivity to the precise definition
of the MS selection box with respect to the adopted stellar evolution models.  For example, Figures 2 and 4 of \citet{mcquinn+11} show that the 
separation between the MS and BHeB sequence narrows towards higher luminosities, 
and that for $M_I \lesssim -7.5$ ($m_I \lesssim 21.3$ ABmag for our observations) 
the MS tilts slightly to the red. Their results are based on stellar evolutionary tracks from the Padova group \citep[specifically from the work of][] {bbcfn1994,gbbc2000,marigo+2008}.  Our Fig.~\ref{fig:cmd_all} shows two stars at $m_I = 21.0$, 21.2 that are outside our selection box, but potentially may be part of a red-ward tilting MS. We note that this red-ward tilt of the bright end of the MS is not discernible in the Padova group evolutionary tracks we adopt.  The addition of two high-luminosity stars to the 789 MS stars already used in our analysis will not significantly alter our results.  However, a wholesale red-ward shift of the red side of our MS selection box, by just a few hundredths of a mag in $B-I$ could add dozens of stars to the analysis.  The ability of our technique to accommodate this contamination would then depend on how well the BHeB sequence can be reproduced by the models.  As noted above, the range of variables that come in to play especially for phases after the MS reduces our confidence of being able to accurately make such a match.

In our analysis we neglect binaries and higher order multiples. Since the stellar luminosity to mass relation is steep for MS stars, then we underestimate the stellar mass and total number of stars if that mass is split in to unresolved multiple stars, compared to our assumption that the star is single. Hence, our assumption will bias the IMF slope to shallow values and the upper-mass limit to higher values compared to reality.  Crowding, if inaccurately corrected, could result in a similar bias.  However, our team applied the same CMD analysis techniques to the optically bright centre of the dwarf galaxy NGC~3741 (at $D = 3.2$ Mpc, similar to M83) and found a best fit IMF much more rich in ionising stars $\alpha \gtrsim -1.9$ \citep{watts_thesis_17}. Hence, our methods are capable of recovering an IMF even more rich in ionising stars than we find in the low surface brightness outer disk of M83.

If some of the sight lines to a group of O stars have a low column density of 
atomic hydrogen, the surrounding \HII\ region is ``leaky''.  Escaping 
ionising photons could either leave the galaxy entirely or create diffuse 
ionised gas well away from the \HII\ region.  In either case, the \Ha\ 
flux of the \HII\ region would underestimate its O star content. The importance 
of the diffuse emission is not clear from the literature.  While some studies 
find diffuse emission contributes up to $\sim60$\%\ of \Ha\ light in nearby 
galaxies, including in their outskirts \citep[e.g.][]{fwgh96,oey+07}, 
\citet{lvmh16} found only $\sim$5\%\ more \Ha\ light in very deep \Ha\ 
imaging of low luminosity dwarf irregular galaxies than they could detect in 
exposures taken at depths more typically found in the literature. The galaxies 
in their sample may be considered analogous to the outer disk of M83 in that 
they have weak \Ha\ emission compared to that in the FUV. Direct detection 
of leaking ionizing flux have been made for a few intensely star forming 
galaxies at $z \sim 0.04$ \citep{bzaamo06,leitet+13,lhlo16} yielding absolute 
escape fractions $< 10$\%. Less ionizing light 
is expected to get out of the disks of normal galaxies. \citet{bhm99} 
estimate an of ionizing photon leakage fraction of a few percent from the 
(presumably typical) disk of the Milky Way and into the halo.  

There are few constraints on the fraction of ionising light that escapes from 
the outer disks of galaxies. \citet{herawjs13} 
argue that much of the ionizing output of O stars should escape from galaxy 
disks where the \HI\ mass density is less than about 4 M$_\odot$ pc$^{-2}$ 
\citep[typical of the \HI\ column density in our fields][]{heald+16}. However, 
the bright optical emission lines of outer disk \HII\ regions in our W3 field 
and in the XUV disk of NGC~4625 by \citet{GildePaz:2007ij} have flux ratios 
that are best fit by photo-ionisation models of single main-sequence stars in 
the mass range 20 to 40 $M_\odot$ producing a normal ionisation bound \HII\ region, 
and are poorly fit with models where the ionising source is a young star 
cluster in a density bounded (i.e.\ leaky) \HII\ region. 

We can 
use observations of \HII\ regions in nearby galaxies to make a crude estimate of 
how much ionising light may escape the confines of the \HII\ regions in the 
outer disk of M83.  In nearby galaxies low-luminosity \HII\ regions are found to 
be less leaky than those with high-luminosity, with those having ${\rm 
\log(L_{H\alpha}\, [erg\, s^{\,-1}]}) < 38.9$ having the characteristics of being 
close to ionisation bounded \citep[i.e.\ not leaking ionising 
photons,][]{brzwk00,zbrr02}.  \citet{Pellegrini:2012ii} found that \HII\ regions 
with ${\rm \log(L_{H\alpha}\, [erg\, s^{\,-1}]) < 38}$ contribute only 10-30\%\ of 
cumulative leaked luminosity to the Magellanic Clouds.
The maximum luminosity of the outer disk regions in this study belongs
to \HII-5 with ${\rm log(L_{H\alpha}\, [erg\, s^{−-1}]) = 37.83}$. At
this luminosity, \citet{Pellegrini:2012ii} found about 50\%\ of \HII\
regions in the Magellanic Clouds are optically thin. Among all leaky
regions, the typical leakage fraction was 0.40. Adopting these figures
for a simple Monte Carlo analysis, in our study we would naively
expect to miss about 20\%\ $\pm$ 9\%\ of the true ionising luminosity,
either in the form of DIG below our detection limit or from ionising
photons escaping the galaxy. At this mild level, the observed
$w_{\rm H\alpha}$ could be a lower limit, making the IMF possibly less
deficient in high-mass stars than implied in Section 4.5. 

A much larger fraction of missing ionising photons would be required for a leakage corrected observed $w_{\rm H\alpha}$ to match the $w_{\rm H\alpha}$ values predicted for a Kroupa IMF in our simulations (see Figure~\ref{fig:M83_Ha_compare}). In particular, we find that on average the observed \Ha\ flux across all fields is 23\%\ (29\%\ if leakage corrected) of the expected \Ha\ flux for a Kroupa IMF; i.e. the missing ionising flux would have to be over three times more than what we recover with our \Ha\ measurements.  We consider this unlikely. Instead, the consistency between our MSLF constraint and the independent $w_{\rm H\alpha}$ constraint in terms of the paucity of massive stars (${\rm M_u \gtrsim 25\, M_\odot}$) suggests that our initial \Ha\ analysis​ assumption of recovering all of the ionising photons is close to the truth.

In summary, by using models of single non-rotating stars to model the MSLF we may bias our results towards containing more of the most massive stars than would be expected for the arguably more likely scenario that the M83 XUV disk contains multiple stars and strongly rotating B stars. Hence, the actual IMF may be more deficient than the standard Kroupa IMF in the most massive stars than what we derive. While it is plausible that the outer disk is leaking ionising photons, the constraints we place on the IMF assuming all ionising photons are captured is consistent with our analysis of the MSLF.  In other words, the MSLF analysis does not reveal an excess population of massive stars that might support the finding of low H$\alpha$ compared to the FUV fluxes in M83, and by extension, other XUV disks. 

\subsection{Non-constant star formation history}
\label{sec:non_const_SFH}

\begin{figure}
\centering
\includegraphics[scale=0.45]{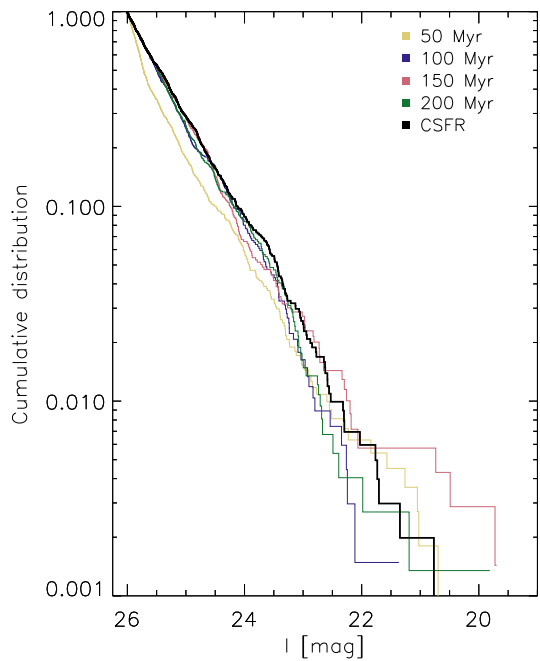} 
\caption{Comparison between the MSLF for different burst scenarios (coloured lines, as labelled) and the MS luminosity function for a model with a constant SFH (black line). All models employ a Kroupa IMF. Here we show four different Gaussian burst models with a FWHM of 1 Myr, $\rm S_{burst}=0.5$ and $\rm t_{burst}=$ 50, 100, 150 and 200 Myr. Each MSLF is shown as the cumulative distribution of MS stars as a function of $I$ band magnitude.\label{fig:burst_comp}}
\end{figure}

So far we have assumed a constant SFH, as justified in Section \ref{sec:choiceSFH}. However, as noted, the IMF and SFH are degenerate; mathematically the MSLF may also be modelled by a non-constant SFH. Here we perform some tests on simulated observations of the MSLF in stellar populations with a varying SFH.  We emphasise that we are not intending to match the CMD of M83 with these particular simulations nor are the adopted SFHs meant to be plausible realisations of the true SFH of the outer disk. Instead the aim of this subsection is to determine the strength required for SFH variations for them to be demonstrably different from a constant SFR population.  In these simulations we adopt a Kroupa IMF, a total MS population matched to that observed in our M83 observations, and likewise, noise characteristics matched to our observations.

We use the same method outlined in Section \ref{sec:sim_CMDs} to produce model CMDs. We model the SFH to have an underlying constant SFR combined with a Gaussian burst. We produce model CMDs in which we vary the input parameters of the Gaussian burst. These parameters are: $\rm S_{burst}$ the burst strength, which we define here as the fraction of star formation that occurs in the burst to the amount of constant star formation over 300 Myr (the time scale of our simulations), and $\rm t_{burst}$, how long ago the burst occurred. We set the full-width at half-maximum of each burst to be 1 Myr and trial three different burst strengths ($\rm S_{burst}=$ 0.5, 0.1 and 0.03) and four different burst times ($\rm t_{burst}=$ 50, 100, 150 and 200 Myr ago). We then compare the MSLF from each of the burst scenarios to the MSLF with a constant SFR using the K--S test, as before. We use the value of $p$ to determine how sensitive the MSLF is to different burst scenarios. Small $p$ (i.e. $< 0.01$) indicate there is a large difference between the MSLF of a model CMD with a constant SFH compared to the bursty SFH. Table \ref{tab:bursts} lists $p$ for the different burst scenarios. We find that the MSLF is sensitive to high strength bursts ($\rm S_{burst} = 0.5$) that occur $\lesssim$ 100 Myr ago. This can be seen as the large difference between the MSLF for  $\rm t_{burst}=50$ Myr ago and the other burst models for $\rm S_{burst}=0.5$ shown in Figure \ref{fig:burst_comp}. This method is also mildly sensitive to intermediate strength bursts $\rm S_{burst}=0.1$ over the same time range and insensitive to small bursts ($\rm S_{burst}=0.03$) over all  burst times (50-200 Myr ago). 

\begin{table}
\centering
\caption{Tested burst scenarios and their corresponding $p$, determined using the K--S test. The Gaussian models have a FWHM of 1 Myr and the exponential models have a burst which is followed by exponential decay with a half-life of 15 Myr. We compare the MSLF for each burst scenario to the MSLF from a model with a constant SFR. All models have a Kroupa IMF. If $\rm p < 0.01$ then we consider the  distributions to be significantly different. \label{tab:bursts}}
\begin{tabular}{c|c|c|c|}
\hline
$\rm S_{burst}$ & $\rm t_{burst}$ (Myr ago) &  Gaussian & Exponential  \\
 & &  $p$ & $p$ \\
\hline \hline
0.5 & 50 &  1.0e-18 & 3.6e-6  \\
0.5 & 100 & 0.23 & 1.4e-2\\
0.5 & 150 & 0.90 &  0.45 \\
0.5 & 200  & 0.37  &  0.91 \\
\hline
0.1 & 50  & 1.8e-3 & 7.3e-2  \\
0.1 & 100 & 0.71 & 0.63 \\
0.1 & 150  & 0.46  & 0.53\\
0.1 & 200  & 0.42 & 0.78 \\
\hline
0.03 & 50  & 0.47 & 0.87 \\
0.03 & 100  & 0.47 & 0.43\\
0.03 & 150  & 0.94 & 0.89\\
0.03 & 200  & 0.38 & 0.58\\
\hline
\end{tabular}
\end{table}

We also test a ``burst-decay'' model whereby the SFR is constant until it has an instantaneous rise to an arbitrary maximum, and then an exponential decay back to the original constant SFR.  For this model we vary $\rm S_{burst}$, the burst strength, and $\rm t_{burst}$, how long ago the burst commenced. We set the exponential half-life to be 15 Myr in all models and then use the K--S test to determine how well the observed MSLF matches each set of parameters. Table \ref{tab:bursts} lists $p$ for each of the scenarios. We find that the MSLF is only sensitive to large bursts ($\rm S_{burst}=0.5$) that occur $\lesssim$150 Myr ago.  We experimented with longer decay timescales and found that very strong bursts $\rm S_{burst}>0.5$ with an  an exponential half-life greater than 75 Myr will appear as a slowly decreasing SFH and be indistinguishable from a constant SFR with our methods.

In summary, our MSLF analysis is sensitive to recent ($\lesssim 50$ Myr ago)
Gaussian bursts of moderate to large strength ($\rm S_{burst}>0.1$) or
fairly recent ($\lesssim 100$ Myr ago) large strength ($\rm S_{burst}>0.5$)
burst-decay scenarios. This method is insensitive to low strength
Gaussian burst scenarios ($\rm S_{burst}=0.03$) and burst-decay
scenarios with $\rm S_{burst}\leq 0.1$. It is also insensitive to bursts
of either type that occurred $\ge 150$ Myr ago. Such low strength or old
bursts do not produce measurable deviations from the MSLF one would find
from a stellar population forming at a continuous rate producing stars
with a Kroupa IMF. We have not considered longer duration (more
realistic) burst timescales in these simple tests. Longer duration
events will dilute their amplitude for a given ${\rm S_{\rm burst}}$
thus undoubtedly making them more difficult to detect.

The observed MSLF is much more deficient in high-luminosity stars than expected if the stars formed at a constant rate with a Kroupa IMF.  If the stars formed with such an IMF there would have to had been a recent drastic decrease in the SFR simultaneously across all fields to match the MSLF results. We argue this is implausible on causality grounds.  The projected separations between the fields is typically 25 arcmin or 34 kpc and the crossing time between fields is $\sim 200$ Myr at orbital velocities of 170 km s$^{-1}$.  To create the observed strong deviations in the MSLF from that expected for continuous star formation requires variations on timescales considerably shorter than this if the Kroupa IMF holds.  However, there is no known physical mechanism that could synchronise the star formation over such large areas on such short timescales.  Given this, we find it unlikely that a stellar population having a Kroupa IMF with a strongly varying SFH could cause the observed MSLF.

\begin{figure}
\centering
\includegraphics[scale=0.45]{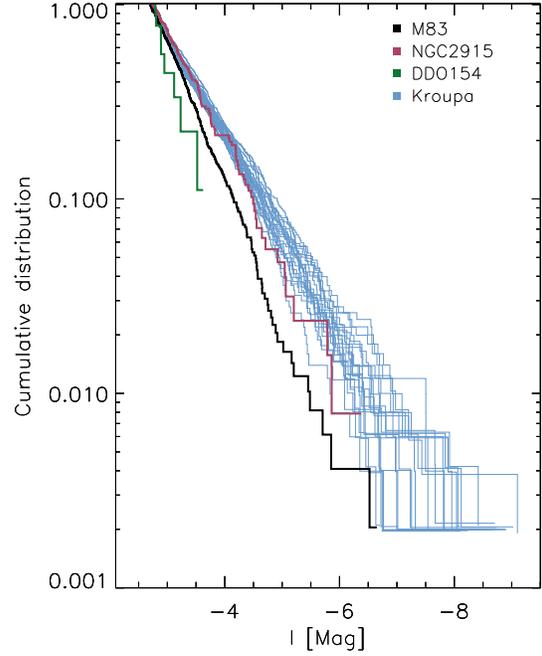} 
\caption{The combined observed MSLF for the outer disk of M83 (black) compared to the observed MSLF in the outer disks of NGC 2915 (brick-red), and DDO~154 (olive-green). The MSLF is shown as the cumulative distribution of MS stars in $I$ band magnitude, down to a limiting $M_I = -2.7$. For comparison, we also show 20 random realisations of a simulated MSLF with a Kroupa IMF (plae-blue) with the same number of stars as the M83 observations to that limit. The cumulative distribution for all the outer disk observations are below the Kroupa IMF realisations especially for high stellar luminosity. M83 is also deficient in high-mass stars compared to NGC 2915, while DDO~154 is the most deficient in high-luminosity stars.  This is in agreement with the best-fitting IMF values for each (see Table \ref{tab:IMF_compare}). \label{fig:MSLF_comp_comb}}
\end{figure}

\subsection{Comparison to other results}
\label{sec:compare}
We compare the observed MSLF in the outer disk of M83 to the previously studied outer disks of NGC 2915 (B15) and DDO~154 \citep{watts+18} in Fig.~\ref{fig:MSLF_comp_comb}. The best-fitting IMF parameters for each outer disk are listed in Table \ref{tab:IMF_compare}. The MS selection box used in each case was tailored to that galaxy.  One important difference is that for both the NGC~2915 and DDO~154 studies the MSLF was originally extracted in the $g$ band, while for M83 the MSLF is extracted in the $I$ band from a selection box in the $I$ versus $B-I$ CMD. For this comparison we use the $I$ MSLF since the different data sets have different detection limits we only consider the stars brighter than a conservative common absolute magnitude limit of $M_I = -2.7$.  This is slightly brighter than our earlier adopted detection limit in the M83 data set in order to be assured that we are well away from the magnitude where crowding limits the number of MS stars in our innermost field.  We list the adopted best fit IMF parameters determined for all three outer disks in Table \ref{tab:IMF_compare}. In Fig.~\ref{fig:MSLF_comp_comb} we also compare the outer disks with 20 random realisations of a Kroupa IMF matched in the number of stars used for the M83 data set. Over most of the plotted range, the cumulative MSLF of NGC 2915 is displaced above the outer disk of M83, but below most of the Kroupa IMF simulations, while DDO~154 has the fewest of the bright stars, with none brighter than $M_I = -4$.  This indicates that all these outer disks are deficient in the highest luminosity stars compared to expectations for a Kroupa IMF, with the deficiency most extreme for DDO~154.

\begin{table}
\centering
\caption{Best-fitting IMF parameters for the outer disks of NGC 2915, DDO~154 and M83. The standard IMF parameters of a Kroupa IMF are included for comparison. \label{tab:IMF_compare}}
\begin{tabular}{c|c|c|l}
\hline
Galaxy    & $\alpha$ & $ \rm M_{u}\, (M_\odot) $ & reference \\ \hline \hline
M83       & -2.35 & 25  & This study \\
NGC 2915  & -2.85 & 60  & B15 \\
DDO 154   & -2.45 & 16  & \citet{watts+18} \\
Kroupa    & -2.35 & 120 & \citet{kroupa01} \\
\hline
\end{tabular}
\end{table}

An independent study of the outer disk of M81 by \citet{gogarten+09} produces results largely consistent with ours.  They use similar tools and techniques to extract $I$ versus $V-I$ CMDs of several different fields: two each centred on \HII\ regions and UV bright regions free of \Ha\ emission.  They interpret their results in terms of the SFH derived from the CMD assuming a constant Salpeter IMF, and find that the \HII\ regions had star formation within the last $\sim$10 Myr, while the UV bright but \Ha\ weak fields had no star formation within the last $\sim$ 16 Myr.  They note that for such recent star formation the diagnostic power of their analysis is limited to the MS, and thus the MSLF.  They show using simulations that the MSLF of the \HII\ regions is different from that of the UV bright - \Ha\ weak fields at the 98\%\ confidence level, with the \HII\ regions producing relatively more of the most luminous stars.  While they also performed some tests where $\alpha$ is varied they did not trial a range of both $M_u$ and $\alpha$ as we do in our models.

Finally, we note that \citet{parker+98} developed a similar technique of constraining the IMF using the MSLF of resolved stellar populations and an assumed SFH. Their study used the UV luminosity function of main-sequence stars observed in the Magellanic Clouds.  They divided their sample between those stars found within \HII\ regions and those found in the ``field'', that is beyond the extent of \Ha\ emission of the HII regions.  They found that that the field star MSLF has a well defined $\alpha = -2.80 \pm 0.09$, whereas $\alpha$ is not as well constrained for the stars within \HII\ regions in which their models allow a broad range of $\alpha$ from $\sim -4.1$ to $\sim -1.9$.  The steepness of the Magellanic Cloud field star IMF is consistent with the results of \citet{cornett+94} and \citet{holtzman+97}, but is not nearly as steep as the value $\alpha = -5.1 \pm 0.2$ found for field stars in the Magellanic Clouds found by \citet{mldg95}.

\section{Conclusions}
\label{sec:concl}
This paper probes the nature of star formation in the extended ultraviolet bright (XUV) disk of M83, concentrating on the young resolved stellar populations seen in HST observations of four fields. Our CMD analysis reveals a clumpy distribution of main-sequence (MS) stars, roughly following the \HI\ distribution and a more smoothly distributed population of red giant branch (RGB) stars. 

We constrain the IMF by comparing the observed main-sequence luminosity function (MSLF) and \Ha\ observations to simulations. We create an extensive ensemble of simulations in which we vary the IMF slope and upper-mass limit. These simulations randomly sample the input IMF and star formation history (SFH) to take into account the stochastic nature of low intensity star formation.  A constant star formation rate (SFR) is adopted as justified by the long dynamical and crossing times in the outer disk. 

The MSLF analysis indicates that an IMF with a power law slope $\alpha=-2.35 \pm 0.3$ and an upper-mass limit $\rm M_{u}=25_{-3}^{+17} M_\odot$ is preferred, i.e.\ a slope matching the Kroupa value, but a lower $M_u$ than the typically quoted value of $\sim100 M_\odot$. There is a degeneracy between the IMF slope and upper-mass limit in our analysis causing large uncertainties in the best-fitting IMF parameters.  However, the best fit region avoids the standard Kroupa IMF parameters. 

To further constrain the form of the IMF, we compare \Ha\ observations to our simulations under the assumption that the fluxes we measure of the \HII\ regions in our fields comprise all the \Ha\ flux in those fields.  A continuously forming stellar population with the same number of MS stars as observed in our fields and standard Kroupa IMF produces about four times more \Ha\ flux than observed.  Deep observations of star forming galaxies and \HII\ regions suggest that more standard observations may miss out on $\sim$ 5\%\ to 25\%\ of the \Ha\ flux \citep{rkelv12,lvmh16}, while the fraction of ionising photons that totally escape a galaxy is expected to be $\lesssim 15$\%\ \citep{dsf00}.  Since low luminosity \HII\ regions, like those seen in the outer disk of M83, have been found to be at best just mildly leaky to ionising photons \citep{brzwk00,zbrr02,Pellegrini:2012ii}, it seems unlikely that escaping ionising photons are able to explain the deficit of \Ha\ flux compared to expectations assuming a standard IMF.  Instead, we find that the observed \Ha\ flux is consistent with the best-fitting IMF parameters from the MSLF analysis. 

Our simulations show that we can discern the difference between star formation at a constant rate and star formation enhanced with a strong and recent Gaussian burst or decaying star formation.  Weak bursts, those older than $\sim 100$ Myr or star formation that has been decaying for $\gtrsim 150$ Myr can not be discerned from star formation at a constant rate with our methods.

Both the MSLF and \Ha\ analysis results indicate that the IMF in the XUV disk of M83 is deficient in high-mass stars compared to a Kroupa IMF. This finding is similar to the results of our previous study of the outer disks of the blue compact dwarf galaxy NGC~2915 (B15) and the dwarf irregular galaxy DDO~154 \citep{watts+18}.  These studies also analysed the MSLF under the assumption of continuous star formation and found the IMF to be deficient in high-mass stars compared to the Kroupa IMF. An IMF deficient in high-mass stars in low density environments, such as the outer disks of galaxies has implications for our interpretation of \Ha\ measurements to measure SFRs, the SFHs derived from CMDs and the chemical evolution of outer disks. Our conclusions challenge the universality of the upper-end IMF. But these results are limited by our inability to constrain the upper-mass limit using optical photometry alone and the small sample size of galaxies analysed with our techniques (three published cases).

There are over 100 galaxies within 5 Mpc \citep[e.g.\ ][]{lee+2011}, many of which 
have been imaged with HST. Hence, there is an opportunity to employ our methods on 
a much larger sample using archival data.  This would allow one to determine how 
the MSLF shape, and presumably the IMF, varies with local parameters (i.e. star 
formation intensity, luminosity, metallicity etc.).  This would put constraints on 
variations in the upper-end IMF in league with constraints on the lower-end IMF 
placed by observations of early type galaxies which have produced mixed results on 
the need to invoke IMF variations \citep[e.g.][]{treu+10,cd12,cappellari+12,slc12,Conroy:2013bt,Ferreras:2013id,asl17,asl18,sle2017,csl2018,parikh+18,sonnenfeld+19,zhou+19,labarbera+19}. In addition, the inclusion of UV photometry would improve upon the optical only 
photometry used in this analysis. Optical photometry alone does not provide 
adequate constraints on both the upper-mass limit and IMF slope; shorter 
wavelengths are needed to differentiate better between different high-mass stars \citep{Bianchi:2006iq,Bianchi:2012he}. 

\section*{Acknowledgements}

We thank the anonymous referee, and the MNRAS editors, for comments which improved the quality of this paper.

This study is based in part on observations made with the NASA/ESA Hubble Space Telescope, obtained at the Space Telescope Science Institute, which is operated by the Association of Universities for Research in Astronomy, under NASA Contract NAS 5-26555. These observations are associated with Program 10608 (PI: D.\ Thilker). Support for Program 10608 was provided by NASA through a grant from the Space Telescope Science Institute. 

This study is based in part on observations made with the Galaxy Evolution Explorer (GALEX).  GALEX is a NASA Small Explorer, launched in 2003 April. We gratefully acknowledge NASA’s support for construction, operation, and science analysis for the GALEX mission, developed in cooperation with the Centre National d'Etudes Spatiales of France and the Korean Ministry of Science and Technology.
 
This study is also based in part on observations at Cerro Tololo Inter-American Observatory, National Optical Astronomy Observatory, which is operated by the Association of Universities for Research in Astronomy (AURA) under a cooperative agreement with the National Science Foundation. 
 
This research has made use of the NASA/IPAC Extragalactic Database (NED) which is operated by the Jet Propulsion Laboratory, California Institute of Technology, under contract with NASA.

SMB was supported by an Australian Postgraduate Award, and this study formed part of her Ph.D.\ thesis \citep{bruzzese_thesis_2016}. CL is funded by a Discovery Early Career Research Award (DE150100618). AGdP acknowledges support from the Spanish MCIUN project AYA2016-75808-R. 

We thank Alessandro Bressan for assistance with the PARSEC stellar evolutionary tracks and George Heald, Frank Bigiel, Ed Elson, and Daniel Dale for providing data and helpful discussions. We thank the GHOSTS team for making their data public. We thank Jin Koda for sending us details on the clusters observed in \citet{koda+12}. We thank Samuel Boissier and Mark Seibert for contributing to our application for the Hubble Space Telescope observations presented here and comments which improved the quality of this paper.  

SMB and GRM thank the Johns Hopkins University for their hospitality and accommodation during several visits while they worked on this project.

\bibliography{m83_wfc_paper}

\renewcommand{\thefigure}{A\arabic{figure}}
\setcounter{figure}{0}

\begin{figure*}
\section*{Appendix A}
\subsection*{M83 HST ACS/WFC images}
\vspace{1cm}
\centering
\includegraphics[width=150mm, height=150mm]{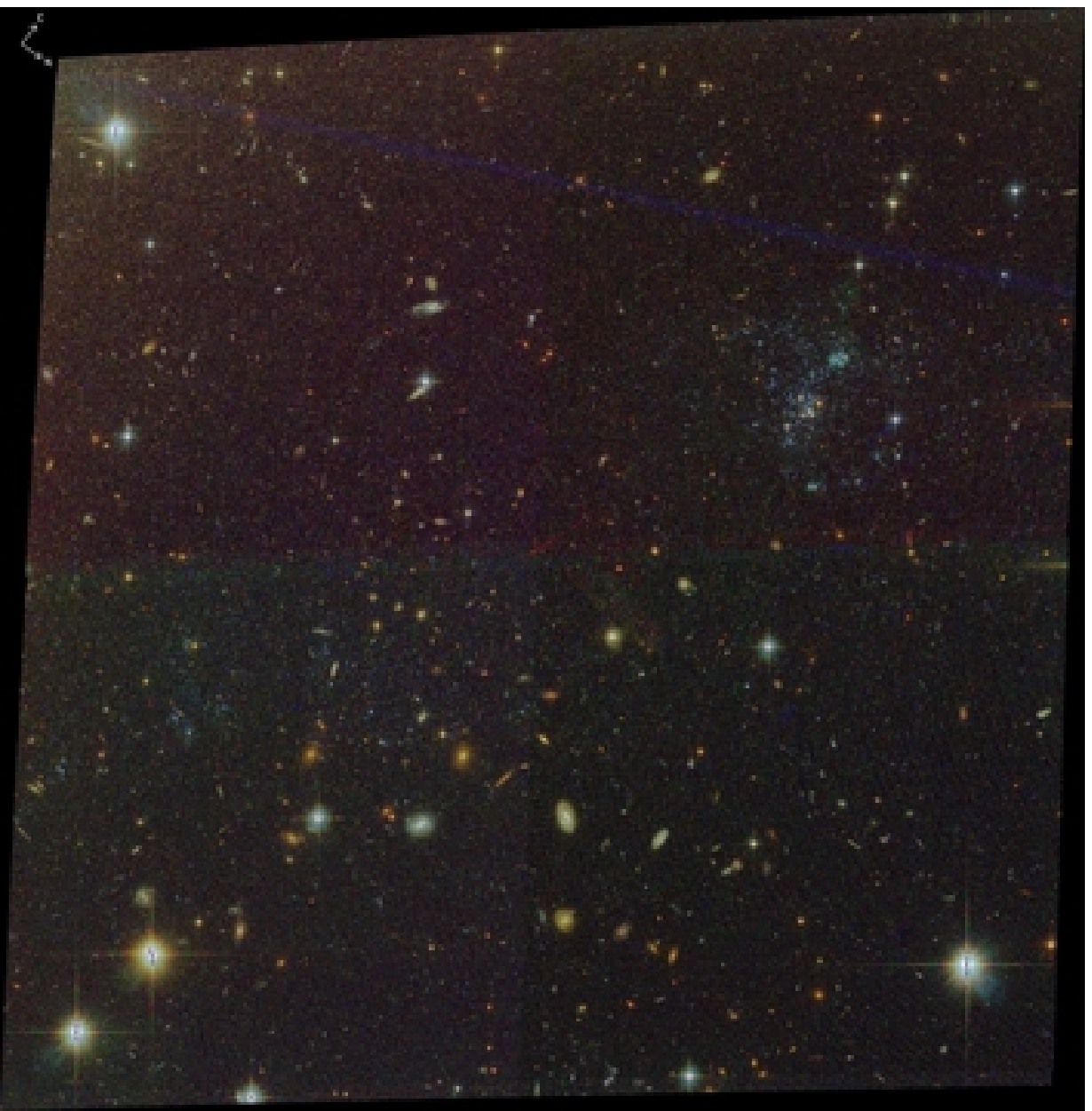}
\caption{Three colour HST ACS/WFC $IVB$ images of field W1.  (See published article for full resolution version of this figure). 
\label{fig:rgb_w1}}
\end{figure*}

\begin{figure*}
\centering
\includegraphics[width=150mm, height=150mm]{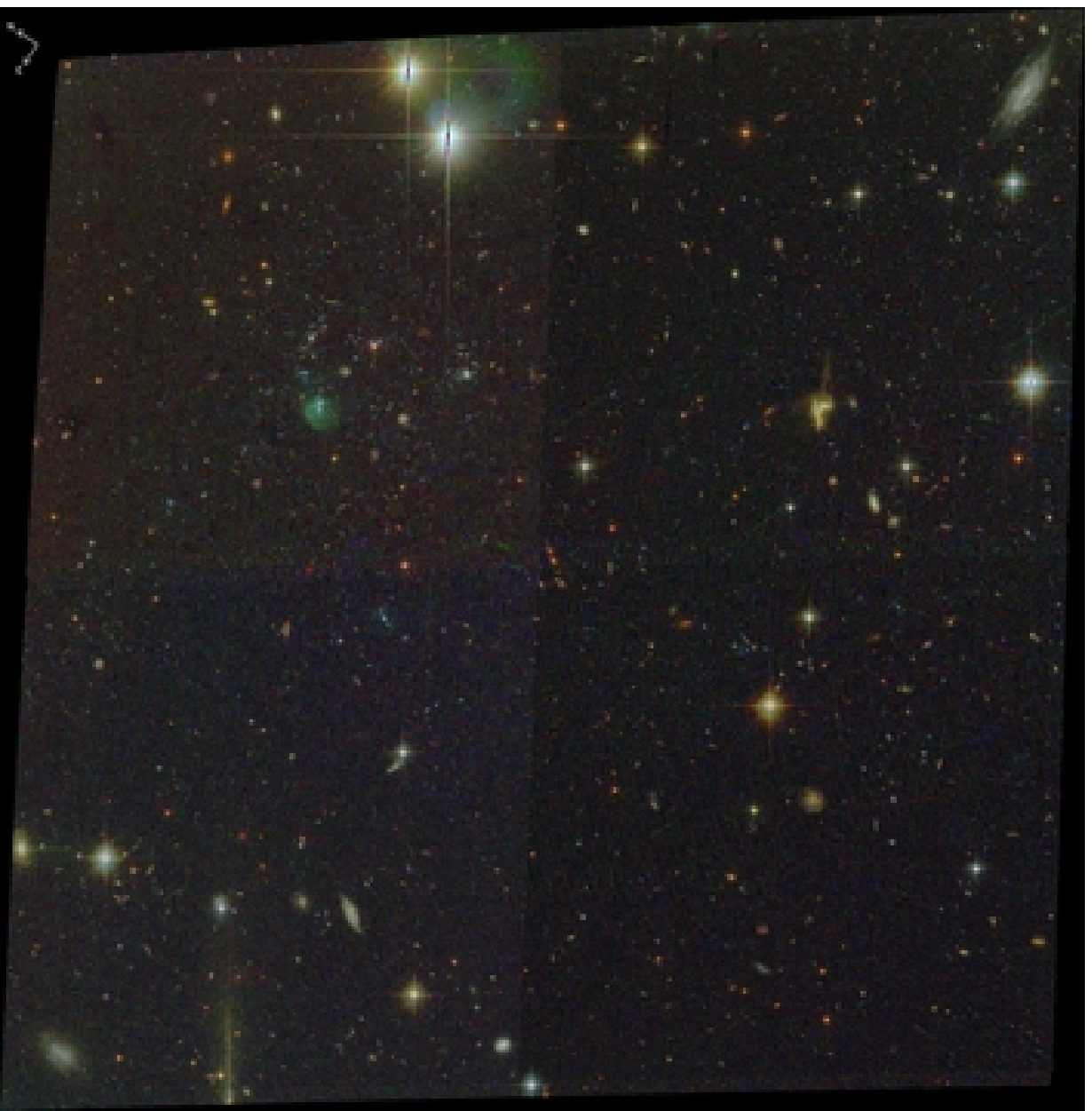}
\caption{Three colour HST ACS/WFC $IVB$ images of field W2.  (See published article for full resolution version of this figure). 
\label{fig:rgb_w2}}
\end{figure*}

\begin{figure*}
\centering
\includegraphics[width=150mm, height=150mm]{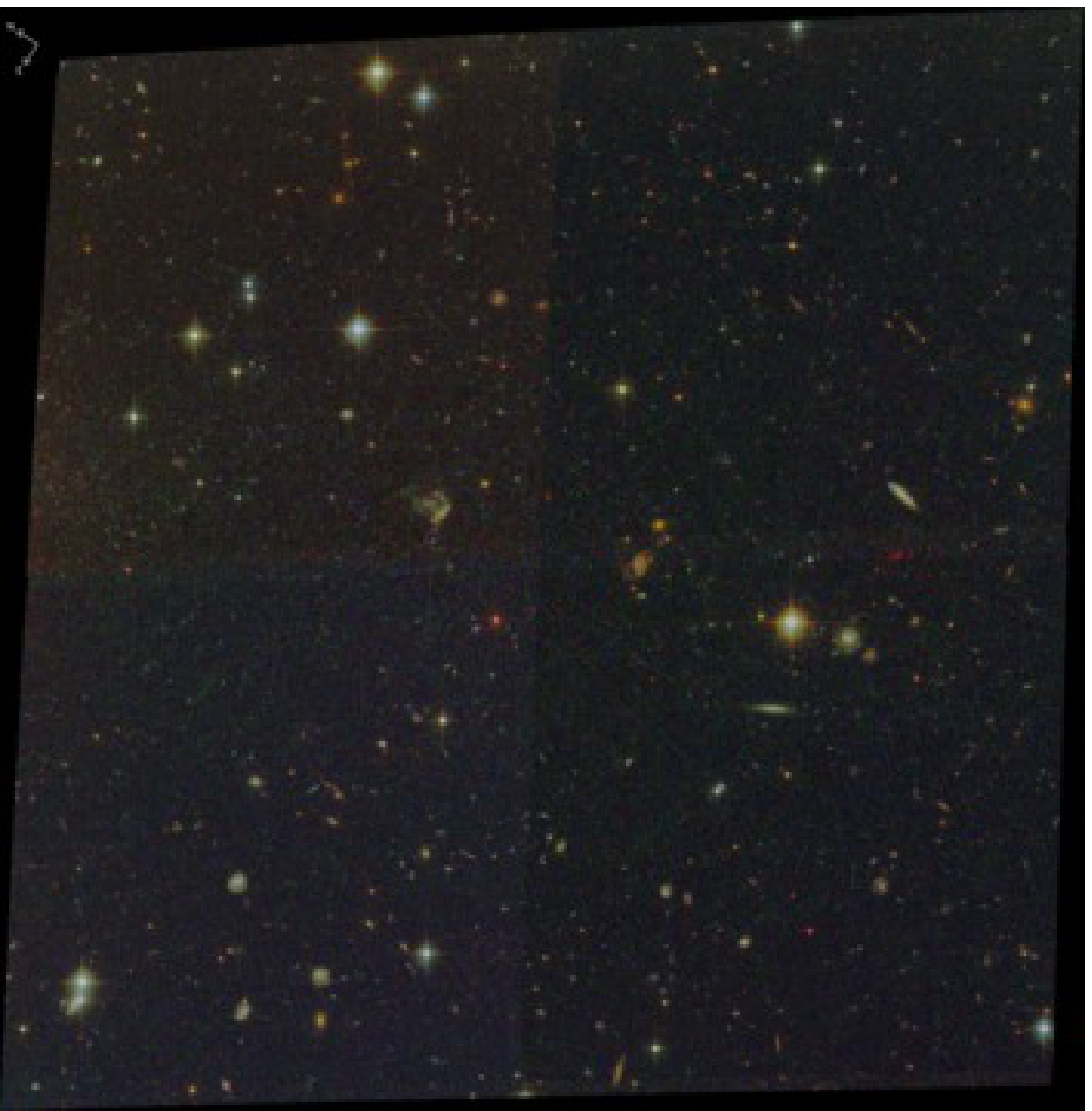}
\caption{Three colour HST ACS/WFC $IVB$ images of field W4.  (See published article for full resolution version of this figure). 
\label{fig:rgb_w4}}
\label{lastpage}
\end{figure*}

\end{document}